%-------------------------
% This paper uses harvmac
%-------------------------
\input harvmac
\noblackbox
%%%%%%%%%%%%%%%%%%%%%%%%%%%%%%%%%%%%%%%%%%%%%%%%%%%%%%%%%%%%%%%%%%%%%%%%%
\def\npb#1#2#3{{\it Nucl.\ Phys.} {\bf B#1} (19#2) #3}
\def\plb#1#2#3{{\it Phys.\ Lett.} {\bf B#1} (19#2) #3}
\def\prl#1#2#3{{\it Phys.\ Rev.\ Lett.} {\bf #1} (19#2) #3}

\def\atmp#1#2#3{{\it Adv.\ Theor.\ Math.\ Phys.} {\bf #1} (19#2) #3}
\def\jhep#1#2#3{{\it JHEP\/} {\bf #1} (19#2) #3}
%%%%%%%%%%%%%%%%%%%%%%%%%%%%%%%%%%%%%%%%%%%%%%%%%%%%%%%%%%%%%%%%%%%%%%%%%
%
%
% titlefont
%
%
\edef\tfontsize{ scaled\magstep3}
 \tfontsize  \tfontsize
 \tfontsize \font\titlei=cmmi10 \tfontsize
\font\titleis=cmmi7 \tfontsize \font\titleiss=cmmi5 \tfontsize
\font\titlesy=cmsy10 \tfontsize \font\titlesys=cmsy7 \tfontsize
\font\titlesyss=cmsy5 \tfontsize  \tfontsize
\skewchar\titlei='177 \skewchar\titleis='177 \skewchar\titleiss='177
\skewchar\titlesy='60 \skewchar\titlesys='60 \skewchar\titlesyss='60
%
%\def\titlefont{\def\rm{\fam0\titlerm}% switch to title font
%\textfont0=\titlerm \scriptfont0=\titlerms \scriptscriptfont0=\titlermss
%\textfont1=\titlei \scriptfont1=\titleis \scriptscriptfont1=\titleiss
%\textfont2=\titlesy \scriptfont2=\titlesys \scriptscriptfont2=\titlesyss
%\textfont\itfam=\titleit \def\it{\fam\itfam\titleit}\rm}
%
%
% math symbols
%
%---------------------------------------------------------------------
%
%
%
% space and backspace in l mode
%
\def\lspace{\ifx\answ\bigans{}\else\qquad\fi}
\def\lbspace{\ifx\answ\bigans{}\else\hskip-.2in\fi} % $$\lbspace...$$
%
%
%     curly letters
%
   %curly letters

%
%
%
%     derivatives
%
%

%

\def\bar#1{\overline{#1}}

\def\abs#1{\left| #1\right|}

\def\half{{\textstyle{1\over2}}} %puts a small half in a displayed eqn
\def\quar{{\textstyle{1\over4}}}
\def\frac#1#2{{\textstyle{#1\over #2}}} %puts a small fraction
%in a displayed eqn
%
%
%     various math operators
%
%

\def\Tr{\mathop{\rm Tr}}

\def\bC{\mathop{\bf C}}

%
%
%
%       relations
%
\def\ltap{\ \raise.3ex\hbox{$<$\kern-.75em\lower1ex\hbox{$\sim$}}\ }
\def\gtap{\ \raise.3ex\hbox{$>$\kern-.75em\lower1ex\hbox{$\sim$}}\ }
\def\gl{\ \raise.5ex\hbox{$>$}\kern-.8em\lower.5ex\hbox{$<$}\ }
\def\roughly#1{\raise.3ex\hbox{$#1$\kern-.75em\lower1ex\hbox{$\sim$}}}
%
%
%       This defines et al., i.e., e.g., cf., etc.

%

\def\prl#1#2#3{{Phys. Rev. Lett. } {#1} (#2) #3}

\def\frac#1#2{{\textstyle{#1 \over #2}}}

\def\[{\left[}
\def\]{\right]}
\def\({\left(}
\def\){\right)}

\def\quar{{1 \over 4}}

\def\frac#1#2{{#1 \over #2}}

\def\semi{\subset\kern-1em\times\;}
\def\bar#1{\overline{#1}}

\def\R{{\bf R}}

\def\phib{\bar{\phi}}

\def\s{\sigma}
\def\t{\tau}

\def\ad{\bar a}

\def\phib{\bar{\phi}}

\def\IR{\relax{\rm I\kern-.18em R}}

\hyphenation{higgs-ino}
%
% more defs
%
\def\hp{\hat{p}}
\def\hq{\hat{q}}
%
% References
%
\lref\weyl{see H.~Weyl, {\it The Theory of Groups and Quantum Mechanics,}
Dover, New York (1950), especially sections 14 and 15 of chapter IV.}
\lref\wigner{E.~Wigner}
\lref\groenewold{H.~Groenewold, Physica {\bf 12} (1946) 405.}
\lref\moyal{J.~Moyal, Proc. Camb. Phil. Soc. {\bf 45} (1949) 99.}
\lref\vonN{J.~von Neumann, Math. Ann. {\bf 104} (1931) 570.}
\lref\david{
J.~R.~David,
``U(1) gauge invariance from open string field theory,''
hep-th/0005085.}
\lref\tatar{R.~Tatar, 
``A Note on Non-Commutative Field Theory and Stability of
Brane-Antibrane Systems,'' hep-th/0009213.}      
\lref\dasgupta{
K.~Dasgupta, S.~Mukhi and G.~Rajesh,
``Noncommutative tachyons,''
JHEP {\bf 0006}, 022 (2000)
[hep-th/0005006].}
\lref\schnabl{M.~Schnabl, ``String field theory at large B-field and
noncommutative geometry,'' hep-th/0010034.}
\lref\kostelecky{V.~A.~Kostelecky and S.~Samuel, 
``On a Nonperturbative Vacuum for the
Open Bosonic String,'' Nucl. Phys. {\bf B336} (1990) 263.}
\lref\sena{A.~Sen and B.~Zwiebach, ``Tachyon Condensation in String
Field Theory,'' hep-th/9912249.}
\lref\senb{A.~Sen, ``Descent Relations Among Bosonic D-branes,'' Int.J. Mod.
Phys. {\bf A14} (1999) 4061, hep-th/9902105.}
\lref\senc{A.~Sen, ``Stable non-BPS bound states of BPS D-branes,''
\jhep{9808}{98}{010}, hep-th/9805019.}
\lref\senoth{A.~Sen,
``SO(32) spinors of type I and other solitons on brane-antibrane pair,''
\jhep{9809}{98}{023}, hep-th/9808141;
``Type I D-particle and its interactions,''
\jhep{9810}{98}{021}, hep-th/9809111; 
``Non-BPS states and branes in string theory,'' 
hep-th/9904207, and references therein.}
\lref\send{A.~Sen, ``BPS D-branes on non-supersymmetric cycles,'' 
\jhep{9812}{98}{021}, hep-th/9812031.}
\lref\bergman{O.~Bergman and M.~R.~Gaberdiel,
``Stable non-BPS D-particles,'' \plb{441}{98}{133}, hep-th/9806155.}
\lref\harveya{J.~A.~Harvey, P.~Horava and P.~Kraus, 
``D-Sphalerons and the Topology
of String Configuration Space,'' hep-th/0001143.}
\lref\berkovits{N.~Berkovits, 
``The Tachyon Potential in Open Neveu-Schwarz String
Field Theory,'' hep-th/0001084.}
\lref\polchinski{J.~Polchinski, ``Dirichlet-Branes and Ramond-Ramond 
Charges,'' \prl{75}{95}{4724}, hep-th/9510017.}
\lref\senspinors{A.~Sen, ``$SO(32)$ Spinors of Type I and Other Solitons on 
Brane-Antibrane Pair,'' \jhep{9809}{98}{023}, hep-th/9808141.}
\lref\phk{P.~Ho\v rava, ``Type IIA D-Branes, K-Theory, and Matrix Theory,'' 
\atmp{2}{99}{1373}, hep-th/9812135.}
\lref\yi{P.~Yi, ``Membranes from Five-Branes and Fundamental Strings from
D$p$-Branes,'' \npb{550}{99}{214}; hep-th/9901159.}
\lref\senpuz{A.~Sen, ``Supersymmetric World-volume Action for Non-BPS
D-branes,'' \hfill\break hep-th/9909062.}
\lref\taylor{W.~Taylor, ``D-brane effective field theory from string
field theory,'' hep-th/0001201.}
\lref\cft{C.~G.~Callan, I.~R.~Klebanov, A.~W.~Ludwig and J.~M.~Maldacena,
``Exact solution of a boundary conformal field theory,'' Nucl.Phys.
{\bf B422} (1994) 417, hep-th/9402113;
J.~Polchinski and L.~Thorlacius, ``Free fermion representation of a boundary
conformal field theory,''Phys. Rev. {\bf D50} (1994) 622, hep-th/9404008;
P.~Fendley, H.~Saleur and N.~P.~Warner, ``Exact solution of a massless
scalar field with a relevant boundary interaction,'' Nucl. Phys.
{\bf B430} (1994) 577, hep-th/9406125;
A.~Recknagel and V.~Schomerus, ``Boundary deformation theory and
moduli spaces of D-branes,'' Nucl. Phys. {\bf B545} (1999) 233,hep-th/9811237.}
\lref\rs{L.~Randall and R.~Sundrum, ``An alternative to compactification,''
Phys.Rev.Lett. {\bf 83} (1999) 4690, hep-th/9906064.}
\lref\agms{M.~Aganagic, R.~Gopakumar, S.~Minwalla and A.~Strominger,
``Unstable solitons in noncommutative gauge theory'',
hep-th/0009142.}
\lref\dbak{D.~Bak,
``Exact multi-vortex solutions in noncommutative Abelian-Higgs theory'',
Phys. Lett. {\bf B495} (2000) 251, hep-th/0008204.}
\lref\jmw{D.~P.~Jatkar, G.~Mandal and S.~R.~Wadia,
``Nielsen-Olesen vortices in noncommutative Abelian Higgs model'',
JHEP {\bf 0009} (2000) 018 , hep-th/0007078.}
\lref\gms{R.~Gopakumar, S.~Minwalla and A.~Strominger,
``Noncommutative solitons'',
JHEP {\bf 0005} (2000) 020 , hep-th/0003160.}
\lref\ci{L.~Cornalba,
``D-brane physics and noncommutative Yang-Mills theory'',
hep-th/9909081; N.~Ishibashi,
``A relation between commutative and noncommutative descriptions of  
D-branes'', hep-th/9909176.}
\lref\witncsft{E.~Witten,
``Noncommutative tachyons and string field theory'',
hep-th/0006071.}
\lref\neksch{N.~Nekrasov and A.~Schwarz,
``Instantons on noncommutative $R^4$ and $(2,0)$ superconformal six  
dimensional theory'',
Commun.\ Math.\ Phys.\  {\bf 198} (1998) 689,
hep-th/9802068.}
\lref\harmor{J.~A.~Harvey and G.~Moore,
``Noncommutative tachyons and K-theory'',
hep-th/0009030.}
\lref\abs{M.~F.~Atiyah, R.~Bott and A.~Shapiro, ``Clifford Modules'',
Topology {\bf 3} suppl. 1 (1964) 3.}
\lref\absapp{E.~Witten,
``D-branes and K-theory'',
JHEP {\bf 9812} (1998) 019, hep-th/9810188;
P.~Horava,
``Type IIA D-branes, K-theory, and matrix theory'',
Adv.\ Theor.\ Math.\ Phys.\  {\bf 2} (1999) 1373,
hep-th/9812135.}
\lref\hklm{J.~A.~Harvey, P.~Kraus, F.~Larsen and E.~J.~Martinec,
``D-branes and strings as non-commutative solitons'',
JHEP {\bf 0007} (2000) 042,
hep-th/0005031.
}
\lref\ekawai{T.~Eguchi and H.~Kawai, ``Reduction of Dynamical Degrees
of Freedom in the Large N Gauge Theory,'' Phys. Rev. Lett. {\bf 48}
(1982) 1063.}
\lref\senunique{A.~Sen, ``Uniqueness of Tachyonic Solitons,''
hep-th/0009090.}
\lref\senissue{A.~Sen, ``Some Issues in Non-commutative Tachyon
Condensation,'' hep-th/0009038.}
\lref\gmsII{R.~Gopakumar, S.~Minwalla and A.~Strominger,
``Symmetry Restoration and Tachyon Condensation in Open String Theory,''
hep-th/0007226.}
\lref\somerefs{
A.~Sen and B.~Zwiebach,
``Large marginal deformations in string field theory,''
JHEP {\bf 0010}, 009 (2000)
[hep-th/0007153];
W.~Taylor,
``Mass generation from tachyon condensation for vector fields on  D-branes,''
JHEP {\bf 0008}, 038 (2000)
[hep-th/0008033];
A.~Iqbal and A.~Naqvi,
``On marginal deformations in superstring field theory,''
hep-th/0008127.}
\lref\soch{C.~Sochichiu, ``Noncommutative Tachyonic Solitons. Interaction
with Gauge Field,'' JHEP {\bf 0008}:026,2000, hep-th/0007217.}
\lref\wittenstrings{E.~Witten, ``Overview of K Theory Applied to Strings,''
hep-th/0007175.}
\lref\nati{N.~Seiberg, ``A Note on Background Independence in Noncommutative
Gauge Theories, Matrix Model and Tachyon Condensation,'' JHEP {\bf 0009}:
003,2000, hep-th/0008013.}
\lref\pioline{B.~Pioline and A.~Schwarz, ``Morita equivalence and T-duality
(or B versus Theta),''
JHEP {\bf 9908} (1999) 021;hep-th/9908019.} 
\lref\truncate{A.~Sen and B.~Zwiebach,
``Tachyon condensation in string field theory'',
JHEP {\bf 0003} (2000) 002,
hep-th/9912249;
N.~Moeller and W.~Taylor,
``Level truncation and the tachyon in open bosonic string field theory'',
Nucl.\ Phys.\  {\bf B583} (2000) 105, hep-th/0002237.}
\lref\tsolrefs{J.~A.~Harvey and P.~Kraus,
``D-branes as Unstable Lumps in Bosonic Open String Field Theory'',
JHEP {\bf 0004} (2000) 012 ,
hep-th/0002117;
N.~Berkovits, A.~Sen and B.~Zwiebach,
``Tachyon condensation in superstring field theory'',
hep-th/0002211;
R.~de Mello Koch, A.~Jevicki, M.~Mihailescu and R.~Tatar,
``Lumps and p-branes in open string field theory'',
Phys.\ Lett.\  {\bf B482}, 249 (2000), hep-th/0003031;
N.~Moeller, A.~Sen and B.~Zwiebach,
``D-branes as tachyon lumps in string field theory'',
JHEP {\bf 0008}, 039 (2000),
hep-th/0005036.}
\lref\gmsii{R.~Gopakumar, S.~Minwalla and A.~Strominger,
``Symmetry restoration and tachyon condensation in open string theory'',
hep-th/0007226.}
\lref\senvac{A.~Sen,
``Some issues in non-commutative tachyon condensation'',
hep-th/0009038; A.~Sen,``Uniqueness of tachyonic solitons'',
hep-th/0009090.}
%\cite{Seiberg:1999vs}:
\lref\sw{
N.~Seiberg and E.~Witten,
``String theory and noncommutative geometry,''
JHEP {\bf 9909}, 032 (1999), hep-th/9908142.}
%%CITATION = HEP-TH 9908142;%%
\lref\bsft{D.~Kutasov, M.~Marino and G.~Moore,
``Some exact results on tachyon condensation in string field theory'',
hep-th/0009148.}
\lref\witcub{E.~Witten,
``Noncommutative Geometry And String Field Theory'',
Nucl.\ Phys.\  {\bf B268} (1986) 253.}
\lref\witbsft{E.~Witten,
``On background independent open string field theory'',
Phys.\ Rev.\  {\bf D46} (1992) 5467, hep-th/9208027.}
\lref\gersh{A.~A.~Gerasimov and S.~L.~Shatashvili,
``On exact tachyon potential in open string field theory'',
JHEP {\bf 0101}:019,2001,
hep-th/0009103.}
\lref\otherbsft{D.~Ghoshal and A.~Sen, ``Normalisation of the 
background independent open string field theory  action'', hep-th/0010021; 
L.~Cornalba,
``Tachyon Condensation in Large Magnetic Fields with Background 
Independent String Field Theory'', hep-th/0009191;
K.~Okuyama,
``Noncommutative Tachyon from Background Independent Open String Field 
Theory'', hep-th/0010028.}
\lref\cds{A.~Connes, M.~Douglas and A.~Schwarz, ``Noncommutative Geometry
and Matrix Theory:Compactification on Tori,'' JHEP {\bf 02} (1998) 003;
hep-th/9711162.}
\lref\schom{V.~Schomerus, ``D-Branes and Deformation Quantization'',
JHEP {\bf 9906} (1999) 030, hep-th/9903205.}
\lref\countb{U.~Venugopalkrishna, ``Fredholm operators associated with strongly
pseudo convex domains in $C^n$,'' J. Functional Anal. {\bf 9}
(1972) 349;      
 L.~Boutet de Monvel, ``On the index of
Toeplitz operators of several complex variables,'' Inv. Math.
{\bf 50} (1979) 249.}
\lref\dh{M.~R.~Douglas and C.~Hull, ``D-branes and the Noncommutative
Torus,'' JHEP {\bf 9802} (1998) 008;hep-th/9711165.}
\lref\furuuchi{K.~Furuuchi,
``Equivalence of projections as gauge equivalence on noncommutative  space'',
hep-th/0005199; K. Furuuchi, ``Topological Charge of $U(1)$ Instantons
on Noncommutative $R^4$'' hep-th/0010006.}
\lref\pho{P.~Ho, ``Twisted bundle on noncommutative space and U(1) instanton'',
hep-th/0003012.}
\lref\grnek{D.~J.~Gross and N.~A.~Nekrasov, ``Monopoles and Strings
in Noncommutative Gauge Theory,'' JHEP {\bf 0007} (2000) 034; hep-th/
0005204;  ``Dynamics of Strings in 
Noncommutative Gauge Theory,'' JHEP {\bf 0010}:021 (2000),  hep-th/0007204.}
\lref\nekrasov{N.~A.~Nekrasov, ``Noncommutative Instantons Revisited,''
hep-th/0010017.}
\lref\poly{A.~P.~Polychronakos,
``Flux tube solutions in noncommutative gauge theories,''
Phys. Lett.{\bf B495} (2000) 407,
hep-th/0007043.}
\lref\gorms{A.~ S.~Gorsky, Y.~M.~Makeeno and K.~G.~Selivanov,
``On Noncommutative Vacua and Noncommutative Solitons,'' 
Phys. Lett. {\bf B492} (2000) 344, hep-th/0007247.}
\lref\hkm{J.~A.~Harvey, D.~Kutasov and E.~J.~Martinec,
``On the relevance of tachyons,''
hep-th/0003101.}
\lref\senvanish{A.~Sen,
``Supersymmetric world-volume action for non-BPS D-branes,''
JHEP {\bf 9910}, 008 (1999)
[hep-th/9909062].}
\lref\dewit{B.~de Wit, J.~Hoppe and H.~Nicolai,
``On the quantum mechanics of supermembranes,''
Nucl.\ Phys.\  {\bf B305}, 545 (1988).}
\lref\coleman{See sec. 3.3 of S.Coleman, ``Quantum lumps and their classical
descendants,'' in {\it Aspects of Symmetry}, Cambridge University Press,
1985.}
\lref\wittenflux{E.~Witten,
``Theta Vacua In Two-Dimensional Quantum Chromodynamics,''
Nuovo Cim.\  {\bf A51}, 325 (1979).}
\lref\bsz{N.~Berkovits, A.~Sen and B.~Zwiebach, 
``Tachyon condensation in superstring field theory,'', 
JHEP {\bf 0003}:002, 2000;
hep-th/0002211.} 
\lref\sena{A.~Sen, ``Stable non-BPS bound states of BPS D-branes,'' 
\jhep{9808}{98}{010}, hep-th/9805019;  
``SO(32) spinors of type I and other solitons on brane-antibrane pair,'' 
\jhep{9809}{98}{023}, hep-th/9808141; 
``Type I D-particle and its interactions,'' 
\jhep{9810}{98}{021}, hep-th/9809111;  
``Non-BPS states and branes in string theory,''  
hep-th/9904207, and references therein.} 
\lref\sennon{A. Sen, ``BPS D-branes on non-supersymmetric cycles,''  
\jhep{9812}{98}{021}, hep-th/9812031.} 
\lref\bergman{O.~Bergman and M.~R.~Gaberdiel, 
``Stable non-BPS D-particles,'' \plb{441}{98}{133}, hep-th/9806155.} 
\lref\senuniv{A.~Sen, 
``Universality of the tachyon potential,'' 
JHEP {\bf 9912} (1999) 027, hep-th/9911116.} 
\lref\polchinski{J.~Polchinski, ``Dirichlet-Branes and Ramond-Ramond  
Charges,'' \prl{75}{95}{4724}, hep-th/9510017.} 
\lref\senspinors{A.~Sen, ``$SO(32)$ Spinors of Type I and Other Solitons on  
Brane-Antibrane Pair,'' \jhep{9809}{98}{023}, hep-th/9808141.} 
\lref\ewk{E. Witten, ``D-Branes and K-Theory,'' \jhep{9812}{98}{019};  
hep-th/9810188.} 
\lref\phk{P. Ho\v rava, ``Type IIA D-Branes, K-Theory, and Matrix Theory,''  
\atmp{2}{99}{1373}, hep-th/9812135.} 
\lref\mm{R. Minasian and G. Moore, ``K-Theory and Ramond-Ramond Charge,''  
\jhep{9711}{97}{002}, hep-th/9710230.} 
\lref\yi{P. Yi, ``Membranes from Five-Branes and Fundamental Strings from 
D$p$-Branes,'' \npb{550}{99}{214}; hep-th/9901159.} 
\lref\senpuz{A. Sen, ``Supersymmetric World-volume Action for Non-BPS 
D-branes,'' \hfill\break 
JHEP {\bf 9910}:008, 1999; hep-th/9909062.} 
\lref\senzw{A. Sen and B. Zwiebach, ``Tachyon Condensation in String Field  
Theory,'' hep-th/9912249.} 
\lref\bcr{M. Bill\'o, B. Craps and F.Roose, ``Ramond-Ramond couplings of 
non-BPS D-branes,'' \jhep{9906}{99}{033}; hep-th/9905157.} 
\lref\sw{N. Seiberg and E. Witten, ``String Theory and Noncommutative  
Geometry,'' JHEP {\bf 9909}:032, 1999; hep-th/9908142.} 
\lref\hk{J. A. Harvey and P. Kraus, ``D-Branes as Lumps in Bosonic 
Open String Field Theory,'' JHEP {\bf 0004}:012,2000, hep-th/0002117.} 
\lref\hkm{J. A. Harvey, D. Kutasov and E. Martinec, ``On the Relevance of 
Tachyons,'' hep-th/0003101.} 
\lref\ks{J. Kogut and L. Susskind ``Vacuum Polarization and the  
Absence of Free Quarks in Four-Dimensions,'' Phys. Rev. {\bf D9} 
(1974) 3501.}  
\lref\cds{A.Connes, M. R. Douglas and A.Schwarz, ``Noncommutative Geometry 
and Matrix Theory:Compactification on Tori,'' JHEP {\bf 9802}, 003 (1998), 
hep-th/9711162.} 
\lref\sendes{A. Sen, ``Descent Relations Among Bosonic D-branes,'' 
Int.J. Mod. Phys. {\bf A14} (1999) 4061, hep-th/9902105.} 
\lref\kjmt{R. de Mello Koch, A. Jevicki, M. Mihailescu and R.Tatar, 
``Lumps and P-Branes in Open String Field Theory,'' hep-th/0003031.} 
\lref\bhy{O.Bergman, K.Hori and P.Yi, ``Confinement on the Brane,'' 
hep-th/0002223.} 
\lref\kosts{V. A. Kostelecky and S.Samuel, ``On a Nonperturbative Vacuum for  
the Open Bosonic String,'' Nucl.Phys. {\bf B336} (1990) 263.} 
\lref\mtaylor{N. Moeller and W. Taylor, ``Level Truncation and 
the Tachyon in Open Bosonic String Field Theory,'' hep-th/0002237.} 
\lref\wtaylor{W. Taylor, ``D-brane Effective Field Theory From String 
Field Theory,'' hep-th/0001201.} 
\lref\hkm{J. A. Harvey, D. Kutasov and E. J. Martinec, 
``On the relevance of tachyons'', hep-th/0003101.} 
\lref\bfss{T. Banks, W. Fischler, S.H. Shenker and L. Susskind, 
``M Theory As A Matrix Model: A Conjecture'',  
Phys.Rev.{\bf D55} (1997) 5112, hep-th/9610043.} 
\lref\dwhn{B.~de Wit, J.~Hoppe and H.~Nicolai, 
``On the quantum mechanics of supermembranes'', 
Nucl.\ Phys.\  {\bf B305} (1988) 545.} 
\lref\garousi{M.~R.~Garousi, 
``Tachyon couplings on non-BPS D-branes and Dirac-Born-Infeld action,'' 
Nucl.Phys. {\bf B584} (2000) 284, hep-th/0003122.} 
\lref\senact{ 
A.~Sen, 
``Supersymmetric world-volume action for non-BPS D-branes,'' 
JHEP {\bf 9910} (1999) 008, 
hep-th/9909062.} 
\lref\bergs{ 
E.~A.~Bergshoeff, M.~de Roo, T.~C.~de Wit, E.~Eyras and S.~Panda, 
``T-duality and actions for non-BPS D-branes,'' 
hep-th/0003221.} 
\lref\kluson{J.~Kluson,
``Proposal for non-BPS D-brane action,''
Phys.\ Rev.\ D {\bf 62}, 126003 (2000),
hep-th/0004106.} 
\lref\dvv{
L.~Motl,
``Proposals on nonperturbative superstring interactions''
hep-th/9701025;
T.~Banks and N.~Seiberg,
``Strings from matrices,''
Nucl.\ Phys.\  {\bf B497} (1997) 41,
hep-th/9702187;
R. Dijkgraaf, E. Verlinde and H. Verlinde, 
``Matrix String Theory'', 
Nucl.Phys. {\bf B500} (1997) 43-61, 
hep-th/9703030.} 
\lref\callm{ 
C.~G.~Callan and J.~Maldacena,  
``Brane Death and Dynamics from the Born-Infeld Action,'' 
Nucl.Phys. {\bf B513} (1998) 198, 
hep-th/9708147.} 
\lref\gibb{G.~W.~Gibbons, 
``Born-Infeld Particles and Dirichlet P-Branes'' 
Nucl.Phys. {\bf B514} (1998) 603, 
hep-th/9709027.} 
\lref\hhk{J.~A.~Harvey, P. Kraus and P. Ho\v rava, 
``D Sphalerons and the Topology of String Configuration Space,'' 
JHEP {\bf 0003}:021,2000, 
hep-th/0001143.}  
\lref\corns{L.Cornalba and R.Schiappa, hep-th/9907211.} 
\lref\ishi{N. Ishibashi, ``A Relation Between Commutative and 
Noncommutative Descriptions of D-Branes,'' hep-th/9909176.} 
\lref\lwone{ 
F.~Larsen and F.~Wilczek, 
``Classical Hair in String Theory I: General Formulation,'' 
Nucl.\ Phys.\  {\bf B475} (1996) 627. 
{hep-th/9604134}.} 
\lref\hlw{P. S.Howe, N.D.Lambert and P.C. West, ``The Selfdual String
Soliton,'' Nucl.Phys. {\bf B515} (1998) 203, hep-th/9709014.}
\lref\dmr{K. Dasgupta, S. Mukhi, and G. Rajesh,
``Noncommutative Tachyons'', hep-th/0005006.}
\lref\malda{J.~Maldacena,
``The large N limit of superconformal field theories and supergravity,''
Adv.\ Theor.\ Math.\ Phys.\  {\bf 2} (1998) 231, hep-th/9711200.}
\lref\luroy{J.X. Lu and S. Roy,
``(p + 1)-Dimensional Noncommutative Yang-Mills and D($p - 2$) Branes'',
hep-th/9912165.}
\lref\sbz{Nathan Berkovits, Ashoke Sen and Barton Zwiebach,
``Tachyon Condensation in Superstring Field Theory'',
hep-th/0002211.}
\lref\wadia{A.~Dhar, G.~Mandal and S.~R.~Wadia,
``String field theory of two-dimensional QCD: 
A Realization of $W_\infty$ algebra,''
Phys.\ Lett.\  {\bf B329} (1994) 15, hep-th/9403050;
``Nonrelativistic fermions, coadjoint orbits of $W_\infty$
and string field theory at $c=1$,''
Mod.\ Phys.\ Lett.\  {\bf A7} (1992) 3129,
hep-th/9207011.}
\lref\reiffel{see M. Rieffel,`` Questions on Quantization,'' quant-ph/9712009
for a brief survey and further references.}
\lref\kons{M. Kontsevich, ``Deformation Quantization of Poisson Manifolds 1.,''
q-alg/9709040.}
\lref\matsuo{Y. Matsuo, ``Topological Charges of Noncommutative Soliton,''
hep-th/0009002.}
\lref\atiyah{M.~Atiyah, {\it K-Theory}, W.~A.~Benjamin, 1967.}
\lref\wegge{N.~E.~Wegge-Olsen, {\it K-Theory and $C^*$-Algebras},
Oxford University Press, 1993.}
\lref\pol{J.~Polchinski,{\it String Theory I,II}, Cambridge University
Press, 1998.}
\lref\hklI{J.~A.~Harvey, P.~Kraus and F.~Larsen,
``Tensionless branes and discrete gauge symmetry,''
Phys.\ Rev.\ {\bf D 63}, 026002 (2001),
hep-th/0008064.}
\lref\hkl{J.~A.~Harvey, P.~Kraus and F.~Larsen,
``Exact noncommutative solitons,''
JHEP{\bf 0012}, 024 (2000),
hep-th/0010060.}
\lref\wong{M.~W.~Wong, {\it Weyl Transforms}, Springer-Verlag, New York, 1998.}
\lref\bdf{L.~G.~ Brown, R.~G.~ Douglas and P.~A.~Fillmore,``Unitary equivalence
modulo the compact operators and extensions of $C^*$-algebras,''
Proc. of a Conference on Operator Theory, pp. 58-128, Lect. Note
in Math. Vol. 345, 1973; ``Extensions of $C^*$-algebras and K-homology,''
Ann. Math. {\bf 105}
(1977) 265.}
\lref\zachos{T.~Curtright, T.~Uematsu and C.~Zachos,``Generating all
Wigner Functions,'' hep-th/0011137.}
\lref\bmoyal{M.~Bartlett and J.~Moyal, Proc. Camb. Phil. Soc. {\bf 45}
(1949) 545.}
\lref\nekra{N.~A.~Nekrasov, ``Trieste Lectures on Solitons in Noncommutative
Gauge Theories,'' hep-th/0011095.}
\lref\komberk{N.~Berkovits, lectures at the Komaba workshop, to appear.}
\lref\komzwie{B.~Zwiebach, lectures at the Komaba workshop, to appear.}
\lref\komtay{W.~Taylor, lectures at the Komaba workshop, to appear.}
\lref\bsft{A.~A.~Gerasimov and S.~L.~Shatashvili,
``On exact tachyon potential in open string field theory,''
JHEP{\bf 0010}, 034 (2000);
hep-th/0009103.}
\lref\kmma{D.~Kutasov, M.~Marino and G.~Moore,
``Some Exact Results on Tachyon Condensation in String Field Theory,''
JHEP {\bf 0010}:045,2000, hep-th/0009148.}
\lref\kmmb{D.~Kutasov, M.~Marino and G.~Moore,
``Remarks on tachyon condensation in superstring field theory,''
hep-th/0010108.}
\lref\cornalba{L.~Cornalba,
``Tachyon condensation in large magnetic fields with background  
independent string field theory,''
hep-th/0010021.}
\lref\okuyama{K.~Okuyama,
``Noncommutative tachyon from background independent open 
string field  theory,''
hep-th/0010028.}
\lref\andreev{O.~Andreev,
``Some computations of partition functions and tachyon 
potentials in  background independent off-shell string theory,''
hep-th/0010218.}
\lref\krausl{P.~Kraus and F.~Larsen,
``Boundary string field theory of the D D-bar system,''
hep-th/0012198.}
\lref\ttu{T.~Takayanagi, S.~Terashima and T.~Uesugi,
``Brane-antibrane action from boundary string field theory,''
hep-th/0012210.}
\lref\tseybsft{A.~A.~Tseytlin,
``Sigma model approach to string theory effective actions with tachyons,''
hep-th/0011033.}
\lref\semirefs{See for example R.~Omnes, 
{\it The Interpretation of Quantum Mechanics,}
Princeton University Press, 1994;
R.~Littlejohn, Phys. Rep. {\bf 138} (1986) 193, and references therein.}
\lref\wittensft{E.~Witten, String Field Theory paper}
\lref\berk{N.~Berkovits,``Super-Poincare Invariant Superstring Field Theory,''
Nucl.Phys. {\bf B450} (1995) 90, hep-th/9503099; ``A New Approach to
Superstring Field Theory,'' proceedings to the 32nd International
Symposium Ahrenshoop on the Theory of Elementary Particles,
Fortschritte der Physik {\bf 48} (2000) 31,hep-th/9912121.}
\lref\konechny{A.~Konechny and A.~Schwarz, ``Introduction to
M(atrix) Theory and Noncommutative Geometry,'' hep-th/0012145.}
\lref\connes{See  A.~Connes, {\it Noncommutative Geometry},
Academic Press 1994.}
\lref\zachrev{C.~Zachos at http://www.hep.anl.gov/czachos/colloq.ps}
\lref\ssta{N.~Seiberg, L.~Susskind and N.~Toumbas, ``Space-Time 
Noncommutatvity and Causality,'' JHEP 0006:044,2000, hep-th/0005015.}
\lref\sstb{N.~Seiberg, L.~Susskind and N.~Toumbas,
``Strings in Background Electric Field, Space/Time Noncommutativity and a
New Noncritical String Theory,'' JHEP 0006:021, 2000, hep-th/0005040.}
\lref\gmms{R.~Gopakumar, J.~Maldacena, S.~Minwalla and A.~Strominger,
``S Duality and Noncommutative Gauge Theory,'' JHEP 0006:036,2000,
hep-th/0005048.}
\lref\grs{O.~J.~Ganor, G.~Rajesh and S.~Sethi, ``Duality and
Noncommutative Gauge Theory,'' Phys. Rev. {\bf D62} (2000) 125008.}
\lref\agw{L.~Alvarez-Gaume and S.~R.~Wadia, ``Gauge Theory on a Quantum Phase
Space,'' hep-th/0006219.}
\lref\djt{G.~V.~Dunne, R.~Jackiw and C.~A.~Trugenberger, ``Topological
(Chern-Simons) Quantum Mechanics,'' Phys. Rev. {\bf D41} (1990) 661.}
\lref\fradtse{E.~S.~Fradkin and A.~A.~Tseytlin, ``Nonlinear Electrodynamics 
From Quantized Strings,'' Phys. Lett. {\bf 163B} (1985) 123.}
\lref\clny{C.~G.~Callan, C.~Lovelace, C.~R.~Nappi, S.~A.~Yost, 
``String Loop Corrections to Beta Functions,'' Nucl. Phys. {\bf B288}
(1987) 525.}
\lref\acny{A.~Abouelsaood, C.~G.~Callan, C.~R.~Nappi and S.~A.~Yost,
``Open Strings in Background Gauge Fields,'' Nucl. Phys. {\bf B280}
(1987) 599.}
\lref\cheung{Y.-K.~E.~Cheung and M.~Krogh, ``Noncommutative Geometry from
0-Branes in a Background B Field,'' Nucl. Phys. {\bf B528}
(1998) 185.}
\lref\chuho{C.-S.~Chu and P.-M.~Ho, ``Noncommutative Open String and
D-Brane, '' Nucl. Phys. {\bf B550} (1999) 151, hep-th/9812219;
``Constrained quantization of open string in background B-field and
noncommutative D-brane,'' Nucl. Phys. {\bf B568} (2000) 447, hep-th/9906192.}
\lref\sheikh{F.~Ardalan, H.~Arfaei and M.~M.~Sheikh-Jabbari, ``Noncommutative
Geometry From Strings and Branes,'' JHEP {\bf 02} 016 (1999), hep-th/9810072;
``Dirac Quantization of Open Strings and Noncommutativity in Branes,''
Nucl. Phys. {\bf B576} (2000) 578,
hep-th/9906161.}
\lref\kadison{R.~V.Kadison and J.~R.~Ringrose, {\it Fundamentals of
the Theory of Operator Algebras}, Graduate Studies in Mathematics, 
Volume 16, American Mathematical Society, 1997.}
\lref\wad{A.~Dhar, G.~Mandal and S.~R.~Wadia, ``String field theory of
two-dimensional QCD: A Realization of $W_\infty$ algebra, '' Phys. Lett.
{\bf B329} (1994) 15, hep-th/9403050; ``Nonrelativistic fermions, coadjoint
orbits of $W_\infty$ and string field theory at $c=1$,''
Mod. Phys. Lett. {\bf A7} (1992) 3129, hep-th/9207011.}
\lref\shatbsft{S.~L.~Shatashvili,
``Comment on the background independent open string theory,''
Phys.\ Lett.\ {\bf B311}, 83 (1993);
hep-th/9303143;
``On the problems with background independence in string theory,''
hep-th/9311177.}
\lref\gniii{D.~J.~Gross and N.~A.~Nekrasov,
``Solitons in noncommutative gauge theory,''
hep-th/0010090.}
\lref\msinger{V.~Mathai and I.~M.~Singer, ``Twisted K-Homology Theory,
Twisted Ext-Theory,'' hep-th/0012046.}
\lref\atiyahs{M.~F.~Atiyah and I.~M.~Singer, ``Index Theory for Skew-Adjoint
Fredholm Operators,'' Inst. Hautes tudes Sci. Publ. math. No. 37 (1969) 5.}
\lref\tseng{L.~Tseng,
``Noncommutative solitons and intersecting D-branes,''
hep-th/0101125.}
\lref\bgl{A.~Bergman, O.~J.~Ganor and J.~L.~Karczmarek,
``A note on intersecting and fluctuating solitons in 4D noncommutative  
field theory,''
hep-th/0101095.}
\lref\hamanaka{M.~Hamanaka and S.~Terashima,
``On exact noncommutative BPS solitons,''
hep-th/0010221.}
\lref\hashimoto{K.~Hashimoto,
``Fluxons and exact BPS solitons in non-commutative gauge theory,''
JHEP {\bf 0012}, 023 (2000),
hep-th/0010251.}
\lref\dealwis{S.~P.~de Alwis and A.~T.~Flournoy,
``Some issues in noncommutative solitons as D-branes,''
hep-th/0011223.}
\lref\blackadar{B. Blackadar, 
{\it K-Theory for Operator Algebras,} Springer-Verlag 1986.}
\lref\kls{M.~Kleban, A.~Lawrence and S.~Shenker,
``Closed strings from nothing,''
hep-th/0012081.}
\lref\ghy{G.~Gibbons, K.~Hori, and P.~Yi, ``String fluid from unstable
D-branes,'' hep-th/0009061.}
\lref\senfun{A.~Sen,``Fundamental Strings in Open String Theory at the
Tachyonic Vacuum,'' hep-th/0010240.}
\lref\chsa{C.~G.~Callan, J.~A.~Harvey and A.~Strominger,
``Worldbrane actions for string solitons,''
Nucl.\ Phys.\ {\bf B367}, 60 (1991).}
\lref\chsb{C.~G.~Callan, J.~A.~Harvey and A.~Strominger,
``World sheet approach to heterotic instantons and solitons,''
Nucl.\ Phys.\ {\bf B359}, 611 (1991).}
\lref\chsc{C.~G.~Callan, J.~A.~Harvey and A.~Strominger,
``Supersymmetric string solitons,''
hep-th/9112030.}
\lref\hetsol{A.~Strominger,
``Heterotic Solitons,''
Nucl.\ Phys.\ {\bf B343}, 167 (1990).}
\lref\emilgreg{E.~J.~Martinec and G.~Moore,
``Noncommutative solitons on orbifolds,''
hep-th/0101199.}
\lref\shatstring{A.~A.~Gerasimov and S.~L.~Shatashvili,
``Stringy Higgs mechanism and the fate of open strings,''
JHEP{\bf 0101}, 019 (2001),
hep-th/0011009.}
\lref\kuiper{N.~H.~Kuiper, ``The homotopy type of the unitary group
of Hilbert space,'' Topology {\bf 3} (1965) 19.}
\lref\periwal{V.~Periwal,
``D-brane charges and K-homology,''
JHEP{\bf 0007}, 041 (2000), hep-th/0006223.}
\lref\bars{I.~Bars, H.~Kajiura, Y.~Matsuo and T.~Takayanagi,
``Tachyon condensation on noncommutative torus,''
hep-th/0010101.}

%
%-------------------
% title page
%-------------------
%
\Title{\vbox{\baselineskip12pt
\hbox{hep-th/0102076}
\hbox{EFI-01-05}
}}
{\vbox{\centerline{Komaba Lectures} \centerline{on}
\centerline{Noncommutative Solitons
and D-Branes}
}}
{
\baselineskip=12pt
\bigskip
\centerline{Jeffrey A. Harvey}
\bigskip
\centerline{\sl Enrico Fermi Institute and Department of Physics}
\centerline{\sl University of Chicago}
\centerline{\sl 5640 Ellis Avenue, Chicago, IL 60637 }
\

\bigskip
\medskip
\centerline{\bf Abstract}
These lectures provide  an introduction to noncommutative geometry and
its origins in quantum mechanics and to the construction of solitons in
noncommutative field theory. These ideas are
applied to the construction of D-branes as solitons of the tachyon field
in noncommutative open string theory. A brief discussion is given
of the K-theory classification
of D-brane charge in terms of the K-theory of operator algebras.
Based on lectures
presented at the Komaba 2000 workshop, Nov. 14-16 2000.
\bigskip
\bigskip

}
\Date{2/01}
%\draftmode
\vfil
\eject
%
%----------------------
% Body of Paper
%----------------------
%\centerline{\bf Contents}
\nobreak\medskip{\baselineskip=12pt
\footnotefont\parskip=0pt\catcode`\@=11
\def\leaderfill#1#2{\leaders\hbox to 1em{\hss.\hss}\hfill%
\ifx\answ\bigans#1\else#2\fi}
\noindent {0.} {Introduction} \leaderfill{2}{2} \par
\noindent {1.} {Lecture 1} \leaderfill{2}{3} \par
\noindent \quad{1.1.} {Motivation} \leaderfill{2}{3} \par
\noindent \quad{1.2.} {Noncommutative Geometry} 
\leaderfill{3}{3} \par
\noindent \quad{1.3.} {The Weyl Transform} \leaderfill{4}{5} \par
\noindent \quad{1.4.} {Noncommutative Geometry in Quantum Mechanics} 
\leaderfill{10}{10} \par
\noindent \quad{1.5.} {Noncommutative Field Theory and Solitons} 
\leaderfill{10}{11} \par
\noindent \quad{1.6.} {The Solution Generating Technique} 
\leaderfill{13}{14} \par
\noindent {2.} {Lecture 2} \leaderfill{15}{l7} \par
\noindent \quad{2.1.} {Derivatives and Gauge Fields in Noncommutative
Field Theory} \leaderfill{15}{l7} \par
\noindent \quad{2.2.} {Noncommutative Solitons at Finite $\theta$} 
\leaderfill{18}{20} \par
\noindent \quad{2.3.} {Vortices and Their  Analogs
in Commutative Field Theory} \leaderfill{19}{22} \par
\noindent \quad{2.4.} {The Noncommutative ABS Construction}
\leaderfill{21}{24} \par
\noindent {3.} {Lecture 3} \leaderfill{23}{26} \par
\noindent \quad{3.1.} {String Theory and Noncommutative Geometry} 
\leaderfill{23}{26} \par
\noindent \quad{3.2.} {Bosonic $D$-branes as Noncommutative Solitons} 
\leaderfill{25}{28} \par
\noindent \quad{3.3.} {Type II $D$-branes as Noncommutative Solitons} 
\leaderfill{29}{33} \par
\noindent {4.} {Lecture 4} \leaderfill{31}{36} \par
\noindent \quad{4.1.} {Noncommutative Solitons and K-theory} 
\leaderfill{31}{36} \par
\noindent \quad{4.2.} {$D$-brane Charge and the Index of Toeplitz Operators} 
\leaderfill{34}{39} \par
\noindent \quad{4.3.} {BDF and all that} \leaderfill{36}{41} \par
\catcode`\@=12\bigbreak\bigskip}
\vfil
\eject

\secno=-1
\newsec{Introduction}

These notes are based on lectures given at the Komaba workshop in November
of 2000. The notes follow the lectures fairly closely, although I have
tried to correct minor errors made in the lectures and to expand on some
points which may have caused confusion.  In addition, I have divided
the third lecture into two lectures  and have included
more material on the relation between $D$-branes and the  $K$-theory of 
operator algebras than
I had  time to present in the original lectures.

The first lecture
discusses how noncommutative geometry arises in quantum mechanics,
formulates simple noncommutative field theories, and constructs
soliton solutions of these theories. In the second lecture derivatives
and gauge fields are introduced and the soliton solutions are
generalized in various ways. The third lecture summarizes how noncommutative
field theory arises in string theory, discusses unstable $D$-brane
systems and their tachyon excitations, and shows that the noncommutative
solitons constructed in the previous lecture lead to configurations
of the tachyon field that  have the properties
of $D$-branes. The final lecture discusses more mathematical material
(although at a fairly low-brow level) which shows how viewing
$D$-branes as noncommutative solitons naturally fits into the
rich theory of operator algebras and their K-theory.  

Many aspects of 
noncommutative field theory and tachyon condensation in string theory
are not covered here. More mathematically sophisticated 
reviews of solitons in noncommutative gauge
theory which cover many topics not discussed here can found in
\refs{\nekra,\konechny}. These references also provide more
complete references to the mathematical literature. The star product
and its historical origin in quantum mechanics is reviewed in
\zachrev.  The
lectures by Berkovits \komberk, Taylor \komtay, and Zwiebach \komzwie\
at this meeting review much of the recent progress in tachyon condensation
using cubic string field theory \witcub\  and its 
supersymmetric generalization \berk. Another approach to tachyon condensation
not
discussed at this meeting is based on boundary conformal field theory
and the closely related techniques of  background independent string field 
theory \refs{\witbsft,\shatbsft}. See 
\refs{\cft,\hkm,\gersh, \kmma,\kmmb,\cornalba,\okuyama,\andreev,\tseybsft,
\krausl,\ttu} 
for treatments using this approach. 

\newsec{Lecture 1}
\subsec{Motivation}
In these lectures I will be discussing the application of noncommutative
geometry and field theory to the problem of open string tachyon condensation,
specifically to the problem of constructing D-branes as solitons
in the tachyon field. There are various reasons why I think this
topic is of some interest. First, noncommutative geometry and
field theory have an interesting structure which parallels
the structure of string theory. Soliton solutions in field theory and
string theory are always interesting and often shed light on the
non-perturbative and strong coupling behavior of the theory, thus
these solutions should be investigated in noncommutative field
theories. That noncommutative field theory arises in a limit
of string theory makes this pursuit even more compelling.
Second, open string tachyon condensation is one example of a
class of problems  that we need to understand better in string theory, namely
identifying  the  vacuum and its excitations in the absence
of supersymmetry. In addition, if we are able to obtain 
a clear understanding of how closed
string physics emerges after condensation of open string modes, it is
likely to teach us new things about the connection between open and
closed strings.  Finally, the connection between  tachyon condensation and
noncommutative geometry  
is in its own right quite striking and suggests that noncommutative
geometry plays a more fundamental role in string theory than I 
would have previously suspected .

\subsec{Noncommutative Geometry}
Noncommutative geometry is based on the following idea \connes.
The structure of
an ordinary commutative manifold $M$  can be captured algebraically by
the algebra $A=C^\infty(M)$ of smooth functions on $M$, 
$f: M \rightarrow \bC$ with the product in the algebra being the
commutative multiplication of functions.  
Although it is hard to see what noncommutative
geometry should be directly, noncommutative algebras are quite familiar.
Thus one tries to deform the commutative algebra $A$ to a noncommutative
algebra with a product $*_\hbar$ such that
\eqn\oneaa{f *_\hbar g = fg + \hbar P(f,g) + \cdots }
where $P$ is a bilinear map $P: A \times A \rightarrow A$. As will be explained
below, the $\hbar$
notation is supposed to suggest an analogy to quantum mechanics. 
As $\hbar \rightarrow 0$ the noncommutative
algebra approaches the commutative algebra defined by the ordinary
product of functions. Given such a noncommutative algebra, the
noncommutative geometry is then defined in terms of noncommutative
generalizations of  the algebraic 
constructs
corresponding to various elements in geometry.  For example, points
in the commutative theory can be defined in terms of ideals since
functions vanishing at a point form an ideal in the algebra $A$.

We will not pursue this very general level of discussion as
there is to my knowledge no general definition of noncommutative
geometry which starts  with an arbitrary manifold $M$ \reiffel. 
An example which has been much studied in the literature occurs when
$M$ is taken to be the phase space of a classical system (that is $M$
is a Poisson manifold) and
$P(f,g)=\{f,g\}_{P.B.}$ is the Poisson bracket \kons.  I will
specialize even further and take $M=\R^{2n}$ with a star product
defined by
\eqn\onebb{f * g(x) = e^{{i \over 2} \theta^{ij} \partial_i \partial'_j}
f(x)g(x')|_{x^i=x'^i} }
where $\theta^{ij}=-\theta^{ji}$ is a non-degenerate constant antisymmetric 
matrix (thus  $\theta$ defines  a symplectic form and
the skew eigenvalues of $\theta$ are analogous to the deformation parameter
$\hbar$).

Before explaining the origin of the star product, first note
the following.  For small $\theta^{ij}$ we have
\eqn\onecc{f*g(x) = f(x)g(x) + { i \over 4} 
\theta^{ij}(\partial_i f \partial_j g-
\partial_j f \partial_i g ) + \cdots }
so that the order $\theta$ term is indeed the Poisson bracket
with symplectic form $\theta^{ij}$. 
Equation \onebb\ thus  implies the star bracket
\eqn\onedd{x^i * x^j - x^j * x^i \equiv [x^i,x^j] = i \theta^{ij}.}

Above we have taken $f,g$ to be complex functions. Later we will
also want to consider real functions, $f: M \rightarrow \R$. If $f,g$
are real functions then  $f*g$ is not in general real since
$(f*g)^{\rm c.c} = (g*f)$ (we denote complex conjugation by
${\rm c.c}$). However $f*f$
is real if $f$ is,  and we will use
this later to construct an action for a real scalar field on
a noncommutative space. 

The star product takes a simple form in momentum space. Let
\eqn\oneaaa{\tilde f(\vec \tau)=\int d^{2n}x e^{i \vec \tau \cdot \vec x}
f(\vec x)}
be the Fourier transform of $f$ and $\tilde g$ the Fourier transform
of $g$. Then one easily finds that the Fourier transform of $f*g$
is
\eqn\oneaab{\widetilde {f*g}(\vec \tau) = {1 \over (2 \pi)^{2n}}
\int d^{2n} \tau' e^{{i \over 2} \theta^{ij} \tau_i \tau'_j}
\tilde f(\half \vec \tau + \vec \tau') \tilde g(\half \vec \tau -
\vec \tau').}

\subsec{The Weyl Transform}
To explain the origin of the star product in quantum mechanics it
will be useful to first introduce the Weyl transform. 
For simplicity I will take $M=\R^2$ and relabel the coordinates as
$x_1=q,x_2=p$ to conform to standard notation in quantum mechanics.
Then $\theta^{12}= \hbar$ although for simplicity I will
usually  set $\hbar=1$.

In quantum mechanics the coordinate and momentum operators $\hat q, \hat p$
obey the Heisenberg commutation relation,
\eqn\onea{[\hq,\hp] \equiv \hq \hp - \hp \hq = i.}
Weyl \weyl\  proposed that we regard this as the Lie algebra of a group
(now termed the Weyl-Heisenberg  group) with elements
\eqn\oneb{U(\tau, \sigma)= \exp[-i ( \tau \hq + \sigma \hp)].}
If we denote the adjoint operator by $\overline U$ then we have
\eqn\onebbb{\eqalign{ U(\tau,\sigma) \hq~ \overline U(\tau,\sigma) & =
                       \hq - \sigma ,\cr
                   U(\tau,\sigma) \hp~ \overline U(\tau, \sigma) & = 
                   \hp + \tau. \cr }}
Thus $U$ acts as translations in phase space. However, using the 
Campbell-Baker-Hausdorff formula  we find
\eqn\onec{U(\tau_1,\sigma_1)U(\tau_2,\sigma_2)= 
e^{-i(\tau_1 \sigma_2-\sigma_1 \tau_2)/2} U(\tau_1+\tau_2,
\sigma_1+\sigma_2).}
Thus we have a representation
up to a phase, or a projective representation 
of the Abelian group of translations in phase space.
We can as usual  view this as a true 
representation of a larger group with elements
$U'(\alpha,p,q)=e^{i \alpha} U(p,q)$.  We will see that the phase factor
responsible for the projective representation is at the heart of noncommutative
geometry.

Weyl further suggested that we should view operators
$\hat O(\hq,\hp)$ as sums of group elements or more formally
as elements of the group algebra \foot{A group algebra is the algebra formed
by linear combinations of group elements with the bilinear product in the
algebra given by group multiplication.}
\eqn\oned{\hat O_f(\hq,\hp)= {1 \over (2 \pi)^2} \int d \sigma d \tau
U(\tau,\sigma) \tilde f(\tau,\sigma). }
If we take $\tilde f(\tau,\sigma)$ to be the Fourier transform
of a function on phase space,
\eqn\onee{\tilde f(\tau,\sigma)=\int dq dp e^{i(\tau q + \sigma p)} f(q,p),}
then in the limit $\hbar \rightarrow 0$ so that $[\hq,\hp] \rightarrow 0$,
$\hat O_f$ is simply $f$. Furthermore, the formula \oned\  gives an ordering
prescription (called Weyl ordering) for constructing an operator from a 
classical function on phase
space:
\eqn\onef{\hat O_f(\hq,\hp) = {1 \over (2 \pi)^2} \int d \sigma d \tau
dq dp e^{-i \tau(\hq-q)-i \sigma (\hp-p)} f(q,p). }

We can also invert the map given by \oned, $f(q,p) \rightarrow
\hat O_f(\hq,\hp)$, to associate functions to operators.  It is
conventional to break the symmetry between coordinates and momenta by
evaluating matrix elements of operators between coordinate 
eigenstates $|q \rangle$. Using conventions with
\eqn\oneg{\langle q'|q \rangle = \delta(q-q'), \qquad
\langle p' | p \rangle = 2 \pi \delta(p-p'),}
we can project out $\tilde f$ using
\eqn\projf{{\rm Tr}_{\cal H} U(\tau,\sigma) \bar U(\tau',\sigma')= 2 \pi
\delta(\tau-\tau') \delta(\sigma-\sigma'),}
%
%
%we consider $\langle q+ \sigma'/2| \hat O_f(\hq,\hp)|q-\sigma'/2 \rangle$
%and Fourier transform with respect to $q$ to find the  Fourier transform
%of $f$:
%
%\eqn\oneh{\tilde f(\tau',\sigma')= 2 \pi \int dq e^{i q \tau'}
%\langle q+ \sigma'/2| \hat O_f(\hq,\hp)|q-\sigma'/2 \rangle .}
%
and then perform the Fourier transform to find 
\eqn\onei{f(q,p) = \int d \sigma' e^{-ip \sigma'}
\langle q+ \sigma'/2| \hat O_f(\hq,\hp)|q-\sigma'/2 \rangle.}
We will refer to the formulae \oned\ and \onei\ relating functions
on phase space to operators on Hilbert space as the Weyl transform
and will call functions and operators so related Weyl transforms of
each other (in more mathematical treatments \wong\  $f(q,p)$ is called the
Weyl symbol of $\hat O_f$ but we will ignore the distinction between
functions and symbols here).

If $\hat O = | \psi \rangle \langle \psi|$ is the projection operator
onto a state $|\psi \rangle$ then the phase space function corresponding
to $\hat O$ is
\eqn\onek{f_{\psi}(q,p)=\int d \sigma' e^{-ip \sigma'} \langle q+\sigma'/2|
\psi \rangle \langle \psi|q-\sigma'/2 \rangle = \int d \sigma'
e^{-ip \sigma'} \psi(q+\sigma'/2) \psi^*(q-\sigma'/2). }
This function is usually called the Wigner distribution function of the
state $\psi$. The Wigner distribution function and more generally
the Weyl transform \onei\ have many applications, particularly in
the analysis of semi-classical physics \semirefs. 

One useful identity  relates the integral
of a phase space function to the trace of its Weyl transform:
\eqn\onel{\eqalign{\int dq dp f(q,p) = & \int dq dp d\sigma' e^{-ip \sigma'}
\langle q+\sigma'/2|\hat O_f|q-\sigma'/2\rangle \cr
      = & \int dq d\sigma' 2 \pi \delta(\sigma')
      \langle q+\sigma'/2|\hat O_f|q-\sigma'/2\rangle \cr
      = & 2 \pi \int dq \langle q|\hat O_f|q \rangle = 
      2 \pi {\rm Tr}_{\cal H} \hat O \cr }}

It will be useful  to distinguish certain  classes of
operators and to understand their representation via the
Weyl transform. 
Recall that an operator $O$ is said to be bounded if there is a number
$k$ such that $|| O x || \le k ||x||$ for all $x$ in a Hilbert space
${\cal H}$ (which we always take to be separable). The set of all
bounded operators forms a $C^*$ algebra denoted ${\cal B(\cal H)}$.
The compact operators form a subset of bounded operators consisting of
the norm completion of finite rank operators. There are various equivalent
definitions of compact operators. For example,
an operator $K$ is compact if, for every bounded sequence $(x_n)$ in
${\cal H}$, the sequence $(K x_n)$ contains a convergent
subsequence, or equivalently, if the image under $K$ of the unit ball in Hilbert
space is compact. 

If $O$ is bounded and $K$ compact, then  clearly $KO$ and $OK$ are
compact, so the compact operators form an ideal in the algebra
of bounded operators on ${\cal H}$. In fact, the vector space of
compact operators, ${\cal K (\cal H)}$, is the only norm closed and
two-sided ideal in ${\cal B(\cal H)}$ and thus plays a distinguished role.
In the relation between commutative and noncommutative geometry
${\cal B(\cal H)}$ should be thought of as  analogous to
$C_b(X)$, the set of all bounded continuous functions on a space
$X$, while ${\cal K(\cal H)}$ should be thought of as the analog
of $C_0(X)$, the $C^*$ algebra of continuous functions on $X$ which
vanish at infinity. This interpretation is supported by the following
relation between operators and their Weyl transforms.

Let ${\cal S}^m$ be the set of functions on phase space defined by
the condition
\eqn\smbound{{\cal S}^m = \{ f(q,p)|~ |\partial_q^\alpha 
\partial_p^\beta f| \le C_{\alpha \beta} (1+q^2+p^2)^{(m-\alpha-\beta)/2} \} }
for constants $C_{\alpha \beta}$. 
Then one can  show that $f$ being in ${\cal S}^m$ with
$m \le 0$ implies that $\hat O_f \in {\cal B}({\cal H})$ while
$m <0$ implies that $\hat O_f \in {\cal K}({\cal H})$. Thus we see
that the Weyl transforms of bounded operators are bounded functions
and the Weyl transforms of compact operators vanish at infinity in
$\R^2$. These definitions have obvious generalizations to $\R^{2n}$.

In what follows it will be useful to keep two examples in mind.
Let $|n \rangle$, with $n$ a non-negative integer be an 
orthonormal basis for
a separable Hilbert space ${\cal H}$. A rank $k$ projection operator
such as
\eqn\rankp{P_k = |0 \rangle \langle 0| + |1 \rangle \langle 1| +
\cdots + |k-1 \rangle \langle k-1|}
is an important example of a compact operator. An example of a bounded
operator which will appear later is the shift operator
\eqn\shifty{S = \sum_{n=0}^\infty |n+1 \rangle \langle n| }

Although many of the calculations to be presented later can be done
without making an explicit choice of basis for Hilbert space, it can be
useful in developing physical intuition to choose a basis and to work
out the Wigner functions (and their generalizations) using this  basis.
Thus we consider a basis of the separable Hilbert space
${\cal H}=L^2(R)$  consisting of the simple harmonic
oscillator eigenstates $|n \rangle $, $n=0,1,2, \cdots \infty$.
We write an arbitrary operator on ${\cal H}$ as
\eqn\onem{\hat O = \sum_{n,m=0}^\infty O_{n,m} |n \rangle \langle m|.}
Let $f_{nm}(q,p)$ be the Weyl transform of the 
operator $|n \rangle \langle m|$:
\eqn\onen{f_{nm}(q,p) = \int dy e^{-ipy} \langle q+{y \over 2}|n 
\rangle \langle
m|q-{y \over 2} \rangle.}

These functions can most easily be computed by introducing the generating
function
\eqn\oneo{G(\lambda,\bar \lambda,q,p)= \sum_{n,m} {\lambda^n \over
\sqrt{n!}}{\bar \lambda^m \over \sqrt{m!}} f_{nm}(q,p). }
I will follow the treatment in \zachos\ which also contains references
to the original literature. 
Define $\lambda = -(\sigma+i \tau)/\sqrt{2}$, 
$a=(\hat q+i \hat p)/\sqrt{2}$,
and
note that the coherent state
\eqn\onep{\eqalign{|\lambda \rangle = & U(\tau,-\sigma)|0 \rangle =
e^{\lambda \bar a-\lambda^* a }|0 \rangle  \cr
= & e^{-|\lambda|^2/2} e^{\lambda \bar a} |0 \rangle \cr }}
obeys
\eqn\oneq{| \lambda \rangle \langle \lambda| = U|0 \rangle \langle 0| \bar U =
e^{-|\lambda|^2} \sum_{n,m}  {\lambda^n \over
\sqrt{n!}}{\bar \lambda^m \over \sqrt{m!}} |n \rangle \langle m| . }
We thus have
\eqn\oner{G(\lambda,\bar \lambda, q,p)= e^{|\lambda|^2} \int dy e^{-ipy}
\langle q+y/2| U |0 \rangle \langle 0 | \bar U |q-y/2 \rangle. }
Since $\bar U |q-y/2 \rangle = e^{ i \sigma \tau/2 + i \tau 
( q-y/2)} |q -y/2 + \sigma \rangle $, and 
$\langle q|0\rangle = \psi_0(q)=\pi^{-1/4}e^{-q^2/2}$,
the integral is an elementary Gaussian integral and 
gives
\eqn\ones{G(\lambda, \bar \lambda,q,p)=2e^{|\lambda|^2} e^{-(q+\sigma)^2
-(p+\tau)^2}.}
Note in particular that $f_{00}=2 e^{-q^2-p^2}$. 

An explicit formula
for the other $f_{nm}$ can be derived in terms of the Laguerre polynomials,
\eqn\oness{L_n^\alpha(x)= {x^{-\alpha} e^x \over n!} \left( {d \over dx}
\right)^n (e^{-x} x^{\alpha+n}).}
Introduce polar coordinates in phase space so that
$q+ip=r e^{i \phi}$. Then a little simple algebra allows us to
write the generating function as
\eqn\onessa{\eqalign{G= & 2e^{-r^2} e^{\bar \lambda \sqrt{2}re^{i \phi} -
\bar \lambda
\lambda}~e^{\lambda \sqrt{2}re^{-i \phi}} \cr
 = &  2 e^{-r^2} \sum_{n=0}^\infty
{1 \over n!} \left( \bar \lambda \sqrt{2}r e^{i \phi} \right)^n
\left(1-{\lambda \over \sqrt{2} r}e^{-i \phi} \right)^n
e^{\lambda \sqrt{2} r e^{-i \phi}}. \cr }}
Setting $y=2r^2$ and $k= -y^{-1/2} \lambda e^{-i \phi}$ we have
\eqn\onessb{G= 2 e^{-y/2} \sum_{n=0}^\infty {1 \over n!}
\left( \bar \lambda \sqrt{y} e^{i \phi} \right)^n
(1+k)^n e^{-yk}. }
The payoff for these  manipulations is that we can now
use the identity for Laguerre functions
\eqn\onessc{\sum_{m=0}^\infty L_m^{n-m}(y)k^m = e^{-yk}(1+k)^n,}
to write
\eqn\onessd{G= 2 e^{-y/2} \sum_{n=0}^\infty \sum_{m=0}^\infty
{1 \over n!} \left(\bar \lambda \sqrt{y} e^{i \phi}\right)^n
L_m^{n-m}(y) \left(-y^{-1/2} \lambda e^{-i \phi} \right)^m .}
Reading off the powers of $\bar \lambda$ and $\lambda$ then gives
the final formula for $f_{nm}$:
\eqn\onesse{f_{nm}(r,\phi)=2 e^{-r^2} \sqrt{{n! \over m!}}
(-1)^n (2 r^2)^{{m-n \over 2}} e^{i \phi (m-n)} L_n^{m-n}(2 r^2). }

\subsec{Noncommutative Geometry in Quantum Mechanics}
Let me now explain  how the star product (sometimes called the
Moyal \moyal\ or Weyl-Moyal product, although it seems to have first appeared
explicitly in work of Groenewold \groenewold, see also \vonN) 
arises in quantum mechanics.
Quantum Mechanics is inherently noncommutative due to the noncommutative 
algebra of quantum mechanical operators. Since in general
$\hat O_f \hat O_g \ne \hat O_g \hat O_f$, it is natural to ask
how this is reflected in the composition law of the Weyl transforms
$f,g$. We can address this question by working
out the Weyl transform of  the operator $\hat O_f \hat O_g$.
We expect to find a noncommutative composition law that reduces to the
ordinary commutative product of functions as $\hbar \rightarrow 0$.

Using the previous expressions for the Weyl transform we have
\eqn\onet{\eqalign{\hat O_f \hat O_g = & {1 \over (2 \pi)^4} \int
d\s_1 d\t_1 d\s_2
d\t_2 U(\t_1,\s_1)U(\t_2,\s_2) \tilde f(\t_1,\s_1)
\tilde g(\t_2,\s_2) \cr
= &  {1 \over (2 \pi)^4} \int  d\s_1 d\t_1 d\s_2
d\t_2 U(\t_1+\t_2,\s_1+\s_2) e^{{i \over 2}(\t_1 \s_2 -\s_1 \t_2)}
\tilde f(\t_1,\s_1)
\tilde g(\t_2,\s_2). \cr }}
Now change variables to $\tau_3 = \tau_1+\tau_2, \tau_4=(\t_1-\t_2)/2$
and $\s_3=\s_1+\s_2, \s_4=(\s_1-\s_2)/2$ to find
\eqn\oneu{\eqalign{\hat O_f \hat O_g = & {1 \over (2 \pi)^2}  \int  
d \s_3 d \t_3
U(\t_3,\s_3) 
\Bigl[ {1 \over (2 \pi)^2} \int d \s_4 d \t_4
e^{{i \over 2}(\s_3 \t_4 - \t_3 \s_4)} \cr
 &  \tilde f(\t_4+\t_3/2,\s_4+\s_3/2)
\tilde g(-\t_4+\t_3/2,-\s_4+\s_3/2) \Bigr]. \cr }}
From \oneaab\ we see that  the quantity in square brackets
is just the Fourier transform of $f*g$ for noncommutative $\R^2$ and
with $\theta^{12}=1$. Thus we have with this choice understood,
\eqn\onev{\hat O_f \hat O_g = {1 \over (2 \pi)^2} \int d \s_3 d \t_3
U(\t_3,\s_3) \widetilde {f*g}(\t_3,\s_3)  =\hat O_{f*g}.}
In other words, the Weyl transform  takes operator multiplication
into the star product of functions on phase space.

\subsec{Noncommutative Field Theory and Solitons}
We will now leave the realm of quantum mechanics and use the
noncommutative star product which we found there (generalized as
in \onebb) to formulate noncommutative
field theories. Instead of viewing the noncommutative coordinates as
phase space coordinates as in quantum mechanics, 
 we now will think of them as the
spatial
coordinates of a noncommutative space.
We will here consider only space-space noncommutativity. Including
temporal noncommutativity introduces new complications which are only
resolved in string theory  \refs{\ssta,\gmms,\sstb,\grs}.  
We will also assume
that we have skew diagonalized $\theta$ so that
\eqn\onex{\theta^{ij}= \pmatrix{\pmatrix{0&\theta_1\cr -\theta_1&0\cr} &
                            0 &\ldots&0 \cr
                            0& \pmatrix{0 & \theta_2 \cr
                                       -\theta_2 & 0 \cr}&\ldots& 0 \cr
                            \vdots&\vdots&\ddots&\vdots \cr
                            0&0&\ldots& \pmatrix{0&\theta_n\cr -\theta_n
                                      &0\cr} \cr }}

To define noncommutative field theories we consider fields which are
functions of time, some set of commutative coordinates $y^A$ and
noncommuting coordinates $x^i$. To begin  we consider noncommutative
field theories in $2+1$ dimensions involving real scalar fields 
$\phi(t,x^1,x^2)$ (equivalently, the Weyl transform is a Hermitian operator)
with $[x^1,x^2]=i \theta$. Although
we will soon want to introduce derivatives and also gauge fields, let's
start by considering an action with only potential terms to illustrate
one of the new features that arises in the study of solitons in
noncommutative field theories. We thus consider the action
\eqn\oney{S=\int dt dx_1 dx_2 V_*(\phi)}
where the $*$ subscript on $V$ indicates that products of fields are
evaluated using the star product. We can easily insert the $\theta$ dependence
into previous formulae by noting that the rescaled coordinates 
$\tilde x^i = x^i/\sqrt{\theta}$
obey $[\tilde x^1,\tilde x^2] = i$. 

Assuming that $V$ is polynomial
in $\phi$, we can shift $\phi$ by a constant so
that up to an overall constant $V$ has the form
\eqn\onez{V_*(\phi) = {m^2 \over 2} \phi*\phi + c_1 \phi*\phi*\phi
+ \cdots }
Using the Weyl correspondence we can also write the action in terms
of $\hat \phi$, the Weyl transform of $\phi$ (denoted $\hat O_\phi$ 
previously) as
\eqn\oneza{S =  2 \pi \theta  \int dt {\rm Tr}_{\cal H} V(\hat \phi)}
We will  use the operator formalism and  drop
the hats on operators when there is little chance of confusion. Also,
in $\R^{2n}$  the prefactor of $\theta$
is replaced by ${\rm Pf}(\theta^{ij}) = \prod_\alpha \theta_\alpha$. 

The equation of motion for this rather trivial action is just the
vanishing of the first derivative of $V$, $V'(\phi)=0$. Given 
\onez\  this  takes the form
\eqn\onezb{V'(\phi)= c \phi (\phi-\lambda_1) \cdots (\phi-\lambda_n)=0}
where $\lambda_0=0, \lambda_1, \cdots \lambda_n$ are the critical points of 
$V$ and $c$ is a constant. 
In commutative field theory the only solution of this equation would be
a constant field, $\phi=\lambda_i$ for some $i$, but in the noncommutative
theory we can construct nontrivial solutions. In particular, if
\eqn\onezc{\phi = \lambda_i P}
where $P$ is a projection operator, $P^2=P$, we also get a solution since
$V'(\phi)= \lambda_i P \cdots \lambda_i(P-1)=0$.
More generally, if $P_i$ are a set of orthogonal projection operators
then $\phi= \sum_{i} \lambda_i P_i$ is also a solution. Of course since
$1-P$ is a projection operator if  $P$ is,
$\phi= \lambda_i(1-P)$ is also a solution. Which of these solutions we 
choose depends
on what we are trying to describe. If $\phi=0$ is the global minimum of
$V(\phi)$ then finite sums of the form $\phi= \sum_i \lambda_i P_i$ with
$P_i$ finite rank  will
be finite energy excitations above the vacuum given by $\phi=0$. On the
other hand, if some $\lambda_i= \lambda_*$ is the global minimum of $V(\phi)$
then a solution $\phi = \lambda_* (1-P)$ has finite energy. Clearly we also
obtain  solutions that  describe finite energy excitations above (or below) an
unstable vacuum, although the physical interpretation of these solutions is
less clear. 

The existence of  simple soliton solutions in noncommutative scalar field
theories  was first discovered  by Gopakumar, Minwalla
and Strominger \gms \foot{Similar constructions appeared earlier in the
context of Matrix Theory \wad. }.  There are two important caveats  to keep
in mind. First, the expression in terms of projection operators holds
only as long as we define the fields so that $V(\phi)$ has no linear
term as in \onez. Second, as pointed out in \hklm, for special choices
of potential there can be soliton solutions which are not of the form
\onezc.

To see what these solutions look like we can consider a simple
choice of projection operator such as the projection operator onto the
harmonic oscillator ground state, $|0 \rangle \langle 0|$, and write
it as a function on the noncommutative $\R^2$ using the Weyl transform.
Using \onesse\ this gives
\eqn\onezd{P=|0 \rangle \langle 0| \rightarrow 
2 e^{-(x_1^2 + x_2^2)/\theta}. }
Thus we obtain a localized solution, even in a field theory without
kinetic terms \gms! This is possible because the noncommutativity parameter
$\theta$ introduces a length scale into the problem.

The energy of the solution \onezc\ is easily evaluated using
$\phi^n = \lambda_i^n P$ to give
\eqn\oneze{E= 2 \pi \theta {\rm Tr}_{\cal H} V(\phi)=2 \pi \theta  V(\lambda_i)
{\rm Tr}_{\cal H} P = 2 \pi \theta  n V(\lambda_i)}
where $n=\Tr_{\cal H} P$ is the rank of the projection operator $P$.

\subsec{The Solution Generating Technique}
Before going on to add in derivative terms and couplings to gauge fields,
I want to discuss a more formal way to construct this solution which will
be useful when we come to more complicated situations.  This construction
is analogous to solution generating techniques that have been used in
a wide variety of other contexts.

The Lagrangian we have used so far,
\eqn\onezf{{\cal L}=2 \pi \theta {\rm Tr}_{\cal H} V(\phi),}
is invariant under the group of
unitary transformations on Hilbert space, 
$U(\cal H)$\foot{The group $U({\cal H})$ is distinct from 
$U(\infty)$ defined as
the inductive limit of $U(n)$. For example, $U({\cal H})$ is 
contractible \kuiper\
while $ \pi_{2n-1}(U(\infty))=Z $ for $n$ a positive integer.}. Thus
the transformation
\eqn\onezg{\phi \rightarrow U \phi \bar U, \qquad \bar U U=U \bar U=1}
leaves ${\cal L}$ invariant and takes solutions of the equation of motion
to solutions since
\eqn\onezh{{d V \over d \phi } \rightarrow U {d V \over d \phi} \bar U}
(note that this requires the vanishing of linear terms in $V$ as in \onez,
for further discussion  see  \dealwis\ ).
Once we add derivatives and gauge fields we will interpret $U({\cal H})$
as a gauge symmetry and solutions related by $U({\cal H})$ transformations
as gauge equivalent,  but for now we view $U({\cal H})$ as a global symmetry
of the simple action \onezf.

In showing that solutions of the equation of motion transform into
solutions, that is, that the equations of motion transform covariantly,
it was only necessary to use $\bar U U=1$ since this is all that is required
to show that powers of $\phi$ transform covariantly. In a finite dimensional
Hilbert space $\bar U U=1$ would imply that $U \bar U=1$, but this is not
true in an infinite dimensional Hilbert space. Operators $U$ obeying
$ \bar U U=1$ are called isometries because they preserve the metric or
inner product on Hilbert space:
\eqn\onezi{\langle \chi | \psi \rangle \rightarrow \langle \chi | \bar U U
| \psi \rangle = \langle \chi | \psi \rangle. }
If it is also true that $U \bar U=1$ then $U$ is unitary. 
Thus if we find a non-unitary
isometry it will still map solutions to solutions, but these solutions
will not be related by the global symmetry (or later the gauge symmetry)
of the action.
Note that $U \bar U U \bar U = U \bar U$ so $U \bar U$
is a projection operator if $U$ is an isometry. 

The standard example of  a non-unitary
isometry is the shift operator
\eqn\onezj{S: ~|n\rangle \rightarrow |n+1 \rangle, \qquad S=\sum_{n=0}^\infty
|n+1 \rangle \langle n|.}
Clearly $\bar S S=1$ but $S \bar S = 1-P \equiv 1-|0\rangle \langle 0|$.
Note that ${\rm ker}~S= \{\emptyset\}$ and ${\rm ker}~\bar S=\{|0 \rangle \}$
so that the index of $S$ is $-1$.
More generally, $U=S^n$ is a non-unitary isometry with index $-n$ and
$U \bar U = 1-P_n$ with $P_n= \sum_{k=0}^{n-1} |k \rangle \langle k|$.

To apply the solution generating technique
we start with the trivial constant solution 
$\phi = \lambda_i I$
with $I$ the identity operator. Transforming this with $U=S^n$
we obtain the new solution $\phi = S^n \lambda_i I \bar S^n= \lambda_i
(1-P_n)$. We can thus interpret the solution \onezc\ as the result of
acting with the solution generating transformation $U=S^n$ on the
constant solution $\phi=\lambda_i I$, and if we choose $\lambda_i$ to be
a global minimum of $V$, then this solution will describe a finite
energy excitation above this vacuum. 

A slightly more complicated example of this technique is illustrated
by starting with the action for a complex noncommutative scalar field
$\phi$ (in operator language we drop the assumption $\phi = \bar \phi$).
We take
\eqn\onezk{{\cal L} = 2 \pi \theta  {\rm Tr}_{\cal H} \left[
W(\phi \bar \phi-1) + W(\bar \phi \phi -1) \right] }
and assume that $W$ is stationary at $\bar \phi \phi = \phi \bar \phi =1$.
In later applications $W$ will be a ``mexican hat'' potential with a local
maximum at $\phi=0$  and a ring of minima at $|\phi|=1$.

The Lagrangian 
${\cal L}$ and the equations of motion are now invariant under $U({\cal H})
\times U({\cal H})$ transformations,
\eqn\onezl{ \phi \rightarrow V \phi \bar U, \qquad U,V \in U({\cal H})}
and the equations of motion are covariant as long as $\bar U U=\bar V V=1$.
In this case we can generate new solutions from old using independent
non-unitary isometries acting on the left and right.  
Since $\phi = I$ is a solution of the equations
of motion, $\phi = V \bar U$ will be also if $U,V$ are non-unitary isometries.
Note that $\phi = V \bar U$ implies that
\eqn\onezm{ \phi \bar \phi \phi = \phi. }
In general, an operator obeying \onezm\ is called a partial isometry.
Note  that \onezm\ implies that $\bar \phi \phi$
and $\phi \bar \phi$ are projection operators.  On the subspace  of
${\cal H}$ where $\bar \phi \phi$ has eigenvalue $+1$, $\phi$ acts as
an isometry, hence the name partial isometry. 

Choosing $V=S^m$, $U=S^n$, we conclude from the above discussion that
$\phi = S^m \bar S^n$ is a solution to the equations of motion. Later on
we will interpret this solution as $m$ vortices and $n$ anti-vortices or
as $m$ $D7$-branes and $n$ $\overline{D7}$-branes 
following \refs{\hklm,\witncsft}.

\newsec{Lecture 2}

\subsec{Derivatives and Gauge Fields in Noncommutative Field Theory}
So far we have found non-trivial solutions to  trivial noncommutative
field theories. To make these theories more realistic we need to understand
how to add derivative terms and couplings to gauge fields. For a review
of these topics see \agw. 

The inclusion of derivatives again goes back to the work of Weyl \weyl. 
Returning
to the notation of quantum mechanics for a moment, let
$f(q,p)$ be a function on phase space and consider the operator corresponding
to $\partial f/\partial q$. That is,
\eqn\twoa{\hat O_{\partial_q f}= {1 \over (2 \pi)^2} \int d \sigma d \tau dq dp
{\partial f \over \partial q} U(\tau, \sigma) e^{(i \tau q + i \sigma p)}. }
Assuming as usual that $f$ has sufficiently fast fall off that we can
integrate by parts, we find
\eqn\twob{\hat O_{\partial_q f}= {1 \over (2 \pi)^2} \int d \sigma d \tau dq dp
(-i \tau f(q,p)) U(\tau, \sigma) e^{(i \tau q + i \sigma p)}. }

Now note that 
$U~\hat p~\bar U= \hp + \tau$ implies $[U,\hp]=\tau U$. Applying
the same logic to $\hat O_{\partial_p f}$ we conclude that
\eqn\twoc{\hat O_{\partial_q f} = i [\hp,\hat O_f], 
\qquad \hat O_{\partial_p f} =
-i[\hq,\hat O_f]. }
Weyl then noted that if we define the derivatives of an operator $O$ to be
\eqn\twod{\eqalign{ {\partial \hat O \over \partial q } & = i [\hp,\hat O] \cr
                  {\partial \hat O \over \partial p} & = -i [\hq,\hat O], \cr }}
then
the Heisenberg equations of motion,
\eqn\twoe{\eqalign{i {\partial \hp \over \partial t} & = [\hp,\hat H] \cr
                   i {\partial \hq \over \partial t} & = [\hq,\hat H], \cr}}
take the same form as the classical Hamilton equations,
\eqn\twof{\eqalign{ {dp \over dt} & = - {\partial H \over \partial q} \cr
                    {dq \over dt} & = {\partial H \over \partial p}. \cr }}

In the noncommutative situation this is easily generalized. With
$[x^i,x^j]= i \theta^{ij}$ the natural derivative operator is
\eqn\twog{\partial_i = -i \theta_{ij} [x^j, ~~] \equiv
 -i \theta_{ij} {\rm ad}~ x^j }
with $\theta_{ij}=(\theta^{ij})^{-1}$.
This satisfies the properties we expect of a derivative operator,
namely $[\partial_i,\partial_j]=0$, $\partial_i x^j  = \delta_i^j$
as well as
\eqn\twoh{\eqalign{{\rm Linearity}: \qquad & \partial_i (a O_1 + b O_2)=
a \partial_i O_1 + b \partial_i O_2 \cr
                   {\rm Leibniz}: \qquad & \partial_i(O_1 O_2) =
                   (\partial_i O_1) O_2 + O_1 (\partial_i O_2). \cr }}

It is useful to introduce complex coordinates. In two noncommuting
dimensions we set $z=(x^1 + i x^2)/\sqrt{2}$ and define $a= z/\sqrt{\theta}$
which then obeys the commutation relation of a annihilation operator,
$[a,\bar a]=1$. We then have
\eqn\twoi{\eqalign{\partial \equiv \partial_z = &
 - \theta^{-1/2} {\rm ad}~ \bar a \cr
\bar \partial \equiv \partial_{\bar z} = & \theta^{-1/2} {\rm ad}~a. \cr }}
In $2n$ dimensions we will block diagonalize $\theta^{ij}$ as in \onex\ and
introduce complex coordinates $z_\alpha$, $\alpha= 1,2 \cdots n$ for
each pair of noncommuting coordinates. We then have $\partial_\alpha =
- \theta_\alpha^{-1/2}{\rm ad}~ a_\alpha$.

We can now formulate noncommutative field theories that contain kinetic
terms as well as potential terms. For example
we could consider a scalar field
$\phi(t,x_i)$ in $2+1$ dimensions  with action
\eqn\twoj{S = 2 \pi \theta 
\int dt {\rm Tr}_{\cal H} [ \half(\partial_t \phi \partial_t \phi - 
\partial_i \phi \partial_i \phi)   - V(\phi)].}
Note that the global $U({\cal H})$ symmetry taking $\phi \rightarrow
U \phi \bar U$ is broken by the presence of derivative terms.

We are aiming towards a description of noncommutative theories arising
from $D$-branes and these carry a gauge field. Thus from now on we will
focus exclusively on noncommutative theories with gauge fields and we will
see that one can  naturally gauge the $U({\cal H})$ symmetry
which was present in the absence of derivative terms.

To introduce gauge fields we look for a covariant derivative that transforms
covariantly under $U({\cal H})$ transformations.
Thus if
\eqn\twok{\phi \rightarrow U \phi \bar U,}
we want
\eqn\twol{D_i \phi = \partial_i \phi -i [A_i, \phi] \rightarrow U D_i \phi
\bar U. }
In complex coordinates we have
\eqn\twom{D_\alpha \phi \equiv D_{z_\alpha} \phi = - \theta_{\alpha}^{-1/2}
[\bar a_\alpha + i \theta_\alpha^{1/2} A_\alpha, \phi], }
or defining $C_\alpha = \bar a_\alpha + i \theta_\alpha^{1/2} A_\alpha$
we can write
\eqn\twon{D_\alpha \phi  =
 - \theta_\alpha^{-1/2}[C_\alpha, \phi], \qquad
          \bar D_\alpha \phi  = \theta_\alpha
          [\bar C_\alpha, \phi]. }

Equation \twol\ then requires that $C_\alpha$ transform as $C_\alpha
\rightarrow U C_\alpha \bar U$, or that
\eqn\twoo{A_\alpha \rightarrow U A_\alpha \bar U -i \theta_\alpha^{1/2}
U [\bar a_\alpha, \bar U]. }
We can introduce a covariant field strength in the usual way
as
\eqn\twop{F_\alpha \equiv i F_{z_\alpha, \bar z_\alpha} =
i(\partial_\alpha \bar A_\alpha - \bar \partial_\alpha A_\alpha
-i[A_\alpha, \bar A_\alpha]) = \theta_\alpha^{-1} \left(
[C_\alpha, \bar C_\alpha]+1 \right). }
With these ingredients we can construct an action for the noncommutative
fields $(\phi,A_\mu)$  with $U({\cal H})$
gauge symmetry:
\eqn\twoq{S = 2 \pi {\rm Pf}(\theta^{ij}) {\rm Tr}_{\cal H} \left(- \quar
F_{\mu \nu}F^{\mu \nu}
+ \half D^\mu \phi D_\mu \phi - V(\phi) \right).}

Let me make a few comments about this action and its generalizations:
\item{1.} We have implicitly chosen $\phi$ to transform in the adjoint
representation of $U({\cal H})$. There are other possibilities, for example
we could consider the fundamental representation as well with $\phi
\rightarrow U \phi$. I have focused on the adjoint representation with
$D$-brane applications in mind.
\item{2.} We have constructed what might be called noncommutative $U(1)$
gauge theory. This can be easily generalized to noncommutative $U(N)$ gauge
theory by tensoring with $N \times N$ matrices and generalizing operator
products (or star products) to include matrix multiplication. 
This generalization
will be used below in sec 2.4. 
\item{3.}In the limit $\theta_\alpha \rightarrow \infty$ it is easy to
see by rescaling coordinates that the first two terms in \twoq\ involving
the gauge field strength and covariant derivative terms vanish relative
to the potential term \gms.  Thus in this limit the previous noncommutative
soliton becomes a solution to the theory which includes derivatives and
gauge fields.

\subsec{Noncommutative Solitons at Finite $\theta$}

At first glance it would seem substantially more complicated
to find solutions at finite values of $\theta$ when gauge
fields and derivatives are included, and this is true if one
tries to analyze the equations of motion directly. Results
along these lines can be found in 
\refs{\grnek,\poly,\dbak,\jmw,\soch,\gorms,\agms, \gniii}. 
The solution generating technique provides a unified way of understanding
the structure of many of these solutions  as well as of generating
new kinds of solutions \refs{\hamanaka,\hashimoto,\schnabl,\bgl,\tseng}  
since it depends
only on the symmetry structure of the Lagrangian and not on
its detailed form.  I will illustrate this for a choice of potential
which will arise later in string theory applications of this formalism.

Consider a noncommutative theory in $2+1$ dimensions with action
given by \twoq.
Furthermore take the potential $V(\phi - \phi_*)$ to have a local minimum
at $\phi=\phi_*$ and a local maximum at $\phi=0$. Finally, adjust the
constant term in $V$ so that at the local minimum we have $V(0)=0$.

As before, we start with the ``vacuum'' with $\phi = \phi_* I$ and
$C= \bar a$ (so that the gauge fields $A_\mu$ vanish). This is
clearly a solution to the equations of motion, and as before, the
equations of motion transform covariantly under transformations $U$
which are ``almost'' gauge in that $\bar U U=1$ but $U \bar U \ne 1$.
Under this transformation we have
\eqn\twor{\eqalign{\phi & \rightarrow U \phi_* I \bar U = \phi_* U \bar U \cr
                    C & \rightarrow U \bar a \bar U. \cr }}
Choosing $U$ to be the $ n^{\rm th}$ power of the shift operator, $U=S^n$ then
gives the solution
\eqn\twos{\eqalign{\phi = & \phi_* (I-P_n) \cr
                    C = & S^n \bar a \bar S^n \cr
                    \bar C = & S^n a \bar S^n \cr }}

Computing the field strength we find
\eqn\twot{F = {1 \over \theta} \left( [C, \bar C]+1 \right) =
              {1 \over \theta} \left( S^n[\bar a,a] \bar S^n +1 \right) =
              {P_n \over \theta} }
and the energy of the solution is
\eqn\twou{E = 2 \pi \theta {\rm Tr} \left( {1 \over 2} F^2 + {1 \over \theta}
[C,\phi][\bar C,\phi] + V(\phi-\phi_*) \right) = 2 \pi \theta n \left(
{1 \over 2 \theta^2} + V(-\phi_*) \right). }
Three  comments concerning this solution:

\item{1.}The energy depends on the ``flux'' $n$ and the value of $V$
at the local maximum $\phi=0$ but is independent of the detailed form
of $V$. This would not be the case for a soliton in a commutative theory.
\item{2.}If we take $\phi_*=0$, or equivalently just drop the tachyon field,
then this solution reduces to the ``fluxon'' solutions discussed in
\refs{\dbak,\agms,\gniii} 
\item{3.}This construction can be trivially extended to noncommutative
$\R^{2n}$ and again the energy diverges as $\theta_\alpha \rightarrow 0$. 
This in
accord with the lack of finite energy solutions in commutative gauge
theory with the action \twoq\ for $n \ge 2$. In string theory we will
find noncommutative solitons with energy which remains finite as
$\theta_\alpha \rightarrow 0$ by modifying the action \twoq.

\subsec{Vortices and Their  Analogs in Commutative Field Theory}
Before moving onto string theory I would like to discuss one generalization
of noncommutative solitons involving noncommutative versions of vortices
and their generalizations. Let me first recall some aspects of
these solitons in
commutative field theories.

Consider a commutative field theory  with scalar (Higgs) fields and gauge
fields in $2n+1$ dimensions. We will look for soliton solutions which are
localized in $\R^{2n}$ (equivalently we could look for p-brane solutions
in higher dimensions by taking $\R^{2n}$ to be the space transverse to the
p-brane).  As the radial coordinate $r \rightarrow \infty$ in $\R^{2n}$
the scalar fields must approach their values in vacuum in order to have
finite energy. Thus the scalar fields define a map from the sphere at
infinity, $S^{2n-1}$, into ${\cal M}$, the vacuum manifold of the theory.
If
\eqn\twov{\pi_{2n-1}(\cal M) = {\rm Z}}
then we can construct topologically non-trivial field configurations.

To be concrete take the gauge group to be  $U(N)_1 \times
U(N)_2$ with a Higgs field $\phi$ transforming in the bifundamental
representation, $\phi \sim (N, \bar N)$ so that
$\phi \rightarrow V \phi \bar U$ with $U \in U(N)_1$, $V \in U(N)_2$
under gauge transformations.  This theory arises in the low-energy description
of $N$ $Dp- \overline{Dp}$-branes. For a suitable choice of potential the
Higgs vacuum expectation value will break $U(N)_1 \times U(N)_2$ down
to the diagonal subgroup and the vacuum manifold will be
\eqn\twow{{\cal M} = {U(N)_1 \times U(N)_2 \over U(N)} \sim U(N). }
This theory supports topologically non-trivial field configurations in
$\R^{2n}$ since $\pi_{2n-1}(U(N))=Z$ for sufficiently large $N$. For
$N=n=1$ this solution is the well known vortex solution of the Abelian
Higgs model. For the vortex solution the winding of  $\phi$ 
around the $S^1$ at infinity requires that derivatives of $\phi$
scale at large $r$ like $\partial \phi \sim 1/r$. Finite energy then
requires that $D \phi$ vanish at large $r$ faster than $1/r$, which in
turn requires that the gauge field have an angular component
scaling like $1/r$. Hence the gauge field strength scales like
$1/r^2$. In $\R^2$ this is consistent with finite energy, but
for $n>1$ the gauge field contribution to the energy
scales like
\eqn\twox{\int d^{2n}x F^2 \sim \int r^{2n-1} F^2 dr \sim \int
r^{2n-3} dr,}
and diverges at large $r$ unless $n<2$. Thus in higher dimensions
the field configurations with non-trivial topology will have divergent
energy with a conventional action for the gauge field.

As we will see, this problem with divergent energy is solved in 
noncommutative gauge
theory and also apparently in string theory.  With this application in mind
it is useful to find an explicit construction of a topologically non-trivial
gauge field. This can be done using an elegant construction due to
Atiyah, Bott and Shapiro \abs. The construction uses the gamma matrices of the
transverse rotation group to construct explicit generators of
$\pi_{2n-1}(U(N))$.

The $Spin(2n)$ rotation group of $\R^{2n}$
has two irreducible spinor representations, $S_\pm$, of dimension $2^{n-1}$.
Let $\Gamma_i$ be the gamma matrices mapping $S_+$ to $S_-$,
\eqn\twoy{\Gamma_i: ~ S_+ \rightarrow S_- .}
The usual Dirac gamma matrices $\gamma_i$ can be constructed in terms
of the $\Gamma_i$ and their Hermitian conjugates $\bar \Gamma_i$ as
\eqn\twoz{\gamma_i = \pmatrix{ 0 & \Gamma_i \cr
                               \bar \Gamma_i & 0 \cr } .}
The result of ABS is that $\Gamma_i x^i /|x|$ is a generator of
$\pi_{2n-1}(U(2^{n-1}))$ where $x^i$ are Cartesian coordinates on
$\R^{2n}$.

To take advantage of this construction we can consider $U(N) \times U(N)$
gauge theory in $\R^{2n}$ as before but with the special choice
$N=2^{n-1}$. A Higgs field in the bifundamental representation with non-trivial
topology at infinity is then constructed as
\eqn\twoaa{\phi = f(|x|) {\Gamma_i x^i \over |x|} }
with $f(|x|)$ approaching a constant as $|x| \rightarrow \infty$.

\subsec{The Noncommutative ABS Construction}
We can now generalize this construction to the noncommutative situation
following \harmor. We first generalize the vortex construction above
with $N=n=1$. To do this we start with the action for a complex scalar
$\phi$ given in \onezk\ and generalize it to include derivatives
and gauge fields. We can gauge the $U({\cal H}) \times U({\cal H})$ symmetry
under which $\phi$ transforms as $\phi \rightarrow V \phi \bar U$
by introducing gauge fields $A^+_\mu,A^-_\mu$ and covariant
derivatives
\eqn\twobb{\eqalign{D_\mu \phi = & \partial_\mu \phi -i(A^+_\mu \phi -
\phi A^-_\mu) \cr
D_\mu \bar \phi = & \partial_\mu \bar \phi + i ( \bar \phi A^+_\mu -
A^-_\mu \bar \phi). \cr }}
We then consider the action for $A^\pm_\mu, \phi$ given by
\eqn\twocc{\eqalign{S = 2 \pi \theta \int dt  {\rm Tr}_{\cal H}
 \Bigl[ & - \quar (F_{\mu \nu}^+ )^2
- \quar (F_{\mu \nu}^- )^2 + \quar (D^\mu \phi D_\mu \bar \phi +
D^\mu \bar \phi D_\mu \phi) \cr
      &  - W(\phi \bar \phi-1) - W(\phi \bar \phi -1) \Bigr]. \cr }} 

Working on noncommutative $\R^2$ we can generate a solution starting
with the vacuum configuration $\phi=1, C^-=\bar a, C^+=\bar a$.
Acting with $U=S^n,V=S^m$ generates the solution
\eqn\twodd{\eqalign{\phi & = S^m \bar S^n \cr
                      C^- & = S^n \bar a \bar S^n \cr
                      C^+ & = S^m \bar a \bar S^m. \cr }}
Computing the field strengths we find  $F^- = P_n/\theta$, $F^+=P_m/\theta$,
which implies that the fluxes in the two gauge fields are given by
\eqn\twoee{\eqalign{\int d^2x F^+ =& 2 \pi \theta {\rm Tr}_{\cal H}
                                         F^+ = 2 \pi m \cr
                    \int d^2 x F^- = & 2 \pi \theta {\rm Tr}_{\cal H} F^- =
                                      2 \pi n. \cr }}

In the commutative limit we have the symmetry breaking pattern
$U(1)_+ \times U(1)_- \rightarrow U(1)_{\rm diag}$. The broken $U(1)_{\rm rel}$
has generator $Q_{\rm rel}=Q_+-Q_-$ with $Q_\pm$ the generators of $U(1)_\pm$.
The $U(1)_{\rm rel}$ flux carried by the vortex is thus $2 \pi (m-n)$ showing
that we should interpret the solution as $m$ vortices and $n$ anti-vortices.

To generalize this construction to $\R^{2n}$ we follow the ABS construction
and consider  $U(N)$ noncommutative gauge  theory with $N=2^{n-1}$ in
$R^{2n}$. In terms of operators this means that $\phi$ is a map
from ${\cal H} \otimes S^+$ into ${\cal H} \otimes S^-$:
\eqn\twoff{\phi:~ {\cal H} \otimes S^+ \rightarrow {\cal H} \otimes S^-~.}
We would like to generalize the ABS construction by finding a solution
of noncommutative gauge theory with the above field content which
in the commutative limit reduces to $\phi= f(|x|) \Gamma^i x_i $.

To implement the solution generating technique we need an analog of
the shift operator, that is an operator from ${\cal H} \otimes S^+$
to ${\cal H} \otimes S^-$ which is a non-unitary isometry with index $-1$.
It is not hard to see that
\eqn\twogg{\bar {\cal S} ={1 \over \sqrt{\Gamma_j x^j \bar \Gamma_k x^k}}
 \Gamma_i x^i }
does the trick. Note that $\bar {\cal S} {\cal S}=1$. It is also fairly
straightforward to see that ${\rm dim~ker~ \bar {\cal S}} =1$ and that
\eqn\twohh{ {\cal S} \bar {\cal S} = 1 - P_{\rm ker~ \bar {\cal S}}}
where $ P_{\rm ker~ \bar {\cal S}}$ is the projection operator onto
the kernel of $\bar {\cal S}$. 
The relevant calculation is
\eqn\twojj{\Gamma_i x^i \bar \Gamma_j x^j = \sum_{\alpha=1}^n
2 \theta_{\alpha}(N_\alpha+1/2) - i \Sigma_{ij} \theta^{ij} }
where $\alpha=1, \cdots n$ labels the $n$ complex coordinates $z_\alpha$,
$\theta^{ij}$ is assumed to be block diagonalized as in \onex,
\eqn\twokk{\Sigma_{ij} = {1 \over 4}\left( \Gamma_i \bar \Gamma_j -
\Gamma_j \bar \Gamma_i \right), }
are generators in the $S_-$ spinor  representation
of $SO(2n)$ and $N_\alpha = \bar a_\alpha a_\alpha$ are number
operators. Equation \twojj\ implies that the kernel of $\bar {\cal S}$
consists of the oscillator ground state times the lowest weight
spinor of $SO(2n)$.

Since ${\cal S}$ is a non-unitary isometry, the previous argument goes
through and allows us to construct a solution in the noncommutative theory
using the solution generating technique with $U= \bar {\cal S}$ and $V=1$:
\eqn\twoll{\eqalign{\phi & = {\cal S} \cr
            C_\alpha^+ & = \bar a_\alpha \cr
             C_\alpha^- & = \bar {\cal S} \bar a_\alpha {\cal S}. \cr  }}
Computing the field strength as before one finds $F^+=0$ and
$F^- = P_{\rm ker~ \bar \phi}/{\rm Pf}(\theta^{ij})$. 

As mentioned earlier, the solution has finite energy at finite $\theta^{ij}$.
Also as expected, the energy diverges as $\theta^{ij} \rightarrow 0$.
Note that this solution is
trivially generalized to a multi ABS/anti-ABS configuration by taking
$U=\bar {\cal S}^n$ and $V = \bar {\cal S}^m$.

To summarize what we have done so far, we have found that one can construct
exact soliton solutions in a variety of noncommutative Yang-Mills Higgs
theories using the solution generating technique. The scalar (Higgs) field
are constructed in terms of projection operators and partial isometries.

\newsec{Lecture 3}
\subsec{String Theory and Noncommutative Geometry}
The main point of these lectures is the connection between noncommutative
solitons and $D$-branes in string theory, so it is time to explain what
all of this has to do with string theory and tachyon condensation.
Over the last few years it has been realized that noncommutative geometry
arises as a limit of string theory when one considers $D$-branes in a
background $B$ field \cds. 
It would take too long to go through the full
details here, so I will just make a few comments and refer to
the literature for details.

A standard example of noncommutative geometry arises when we consider
a particle of charge $q$ moving in a plane (with coordinates $x,y$) in the
presence of a magnetic field transverse to the plane ($\vec B = B \hat z$).
If we parallel transport the particle around a closed loop in the plane
then we know that the wave function of the particle acquires a phase
proportional to the magnetic flux through the loop. As an example of
such a loop consider a square with sides of length $a$. If $T_x(a)$,
$T_y(a)$ are the translation operators by  $a$ in the $x,y$
directions then we have
\eqn\threea{T_x(a) T_y(a) T_x^{-1}(a) T_y^{-1}(a) = e^{2 \pi i q B a^2}.}
Thus we have a projective representation of the translation group on
the plane. As in quantum mechanics where the plane was two-dimensional
phase space, this can be used to define a noncommutative product of
functions on the plane. 

Indeed, the Lagrangian for a charged particle of
mass $m$ in a constant magnetic field is
\eqn\threeb{{\cal L} = {m \over 2} \left( {d x \over dt}^2 +
{d y \over dt}^2 \right) + q B x {d y \over dt}.}
As is well known, the spectrum consists of infinitely degenerate
Landau levels with the energy spacing between levels given
by $q B /m$. As $m \rightarrow 0$ this
gap becomes infinite and physics is restricted to the lowest
Landau level. In this limit the Lagrangian reduces to the second term
in \threeb\ and the momentum canonically conjugate to $y$ becomes
\eqn\threec{p_y = {\partial {\cal L} \over \partial (dy/dt)} =qBx.}
Thus  the canonical commutation relations imply
\eqn\threed{ [x,y]=iq/B,}
showing that the coordinates on the plane do not commute. This treatment
was rather naive, but a more complete treatment leads to a similar
conclusion \djt. 

Similarly, in string theory a constant  antisymmetric 
two-form potential $B_{\mu \nu}$ couples
to the string via
\eqn\threee{\int_{\Sigma} B_{\mu \nu} \epsilon^{ab} \partial_a X^\mu
\partial_b X^\nu  = \int_{\partial \Sigma} B_{\mu \nu} X^\mu \partial_t
X^\nu}
where $\Sigma$ is the string world-sheet with boundary $\partial \Sigma$
and $\partial_t$ is the derivative tangential to the boundary. Since
with constant $B$ the interaction is a surface term, it does not change
the dynamics unless the string-world sheet has a boundary, that is unless
there are $D$-branes present. In the presence of $D$-branes, the interaction
\threee\ modifies the boundary conditions obeyed by the string coordinates
$X$ and hence changes the propagator. Seiberg and Witten \sw\ showed, using
previous results on the quantization of open strings \refs{\fradtse,\clny,
\acny} and building on previous work \refs{\cheung,\chuho,\sheikh,\schom}
that the effect of
$B$ could be summarized as follows.

There is a two-form field $\Phi$ whose value depends on how
the theory is defined (that is, how it is regularized). The action for the
spacetime fields associated to the different modes of the string depends
on the combination $\hat F_{\mu \nu} + \Phi_{\mu \nu}$ where $\hat F$ is
the gauge field. For further details of the role played by
$\Phi_{\mu \nu}$ see \nati.
 Assuming that $B$ has only spatial components and making
the special choice $\Phi_{ij} = - B_{ij}$, the changes to the effective
action for the open string modes due to the presence of $B$ are:
\item{1.}In the action use the open string metric $G_{ij}$ defined in
terms of the background closed string metric $g_{ij}$ by
\eqn\threef{G_{ij} = - (2 \pi \alpha')^2 \left(B {1 \over g} B \right)_{ij}}
\item{2.}Change the string coupling constant from its value $g_s$
when $B=0$ to
\eqn\threeg{G_s = g_s~{\rm det}(2 \pi \alpha' B g^{-1})^{1/2}}
\item{3.}Use star products to multiply fields with
\eqn\threeh{\theta^{ij} = \left( {1 \over B} \right)^{ij}. }

In the rest of this lecture we will
consider various unstable or non-BPS $D$-brane configurations in the
bosonic string and type II superstring in the presence of a background
$B$ field. The effective field theory of the tachyon and gauge field
degrees of freedom of the open strings on the $D$-brane will be a 
noncommutative
field theory determined by  the above procedure in terms of the effective
field theory at $B=0$. We will see that the noncommutative solitons studied
in the previous lectures can be identified with lower-dimensional $D$-branes
which arise from tachyon condensation \refs{\dasgupta,\hklm}. We will carry
out the analysis using an effective field theory approach, although
for large $B$ the analysis can also be done directly in string
field theory \witncsft.

\subsec{Bosonic $D$-branes as Noncommutative Solitons}
The first example we will consider is the bosonic string. The bosonic
string contains unstable $Dp$-branes for all $p$, for simplicity we
start by considering a space-filling $D25$-brane.  The open string
field theory describing the excitations of the $D25$-brane involves
an infinite number of ordinary component fields, the tachyon $\phi$,
a massless gauge field $A_\mu$, and an infinite tower of massive states.
In string field theory these fields are organized into a string field
$\Phi$ with a gauge transformation law which when expressed in terms
of the component fields is  complicated and  differs from
the usual gauge transformation laws when interactions are included.
Nonetheless, it should be the case that if we integrate out all fields
other than $\phi, A_\mu$ to obtain an effective action, $S_{\rm eff}(
\phi, A_\mu)$, then this effective action is invariant under
standard gauge transformations since this is the only consistent
way to decouple the negative norm states of $A_\mu$.

In the commutative theory the tachyon field is a singlet under
$U(1)$ gauge transformations, but an analysis of disk diagrams
shows that in the noncommutative theory the tachyon is in the
adjoint representation of the noncommutative $U(1)$ gauge theory \garousi.
Also, while higher derivative terms in the effective action
are  not known precisely, it is known that the leading terms for
constant tachyon and gauge field strength have the
Born-Infeld form \refs{\senuniv,\senvanish,\kluson}.

Thus  integrating out the massive string degrees of
freedom leads to an effective action of the form  (assuming for now
that $B_{ij}=0$)
\eqn\threei{S_{\rm eff}= {c \over g_s} \int d^{26}x 
\left[ - V(\phi-1) \sqrt{-{\rm det}(g +
2 \pi \alpha' F ) }  + \cdots +
{1 \over 2} \sqrt{g} f(\phi-1) \partial^\mu \phi \partial_\mu \phi + \cdots
\right]}
Here $c=T_{25} g_s$ is independent of $g_s$ with $T_{25}$ the $D25$-brane
tension. The zero point and scale of the potential $V(\phi-1)$ have  been
chosen so that there is a local maximum at $\phi=0$ with $V(-1)=1$
which represents the unstable $D25$-brane configuration, and a local
minimum at $\phi=1$ with $V(0)=0$ which is supposed to represent the
closed string vacuum according to the conjecture of Sen \refs{\senb,\senc}.
Finally, the ellipsis  in \threei\ indicate 
higher derivative terms.

We now turn on a background $B$ field. The simplest choice (which is
easily generalized) is to turn on a $B$ field along a $\R^2$ in $\R^{25,1}$,
say $B_{24,25}=b$. I will choose $b<0$ for convenience in later expressions.
Following the prescription described above \threef, \threeg\ we need
the open string metric, coupling constant and noncommutativity parameter
which are
\eqn\threej{\eqalign{G_{\mu \nu} & ={\rm diag}(1,-1,-1,\cdot -1,-(2 \pi \alpha'
b)^2,-(2 \pi \alpha' b)^2) \cr
G_s & = g_s (2 \pi \alpha' |b|) \cr
\theta^{24,25} & \equiv \theta = 1/|b|. \cr }}
Using the operator formalism in the noncommuting
directions the action then becomes
\eqn\threek{S = {2 \pi \theta c \over G_s} \int d^{24}x  
{\cal L}_{\rm nc}}
with
\eqn\threej{{\cal L}_{\rm nc}={\rm Tr}_{\cal H} \left[ -  V(\phi-1)
\sqrt{det(G_{\mu \nu}  2 \pi  \alpha' (F+\Phi)_{\mu \nu})}
+ {1 \over 2}\sqrt{G} f(\phi-1) D^\mu \phi D_\mu \phi + \cdots \right]}
where $F_{24,25}+\Phi_{24,25}=-iF_{z,\bar z}+ 1/\theta = -[C,\bar C]/\theta$.

We can now use the solution generating technique to construct
a solution starting from the vacuum configuration\foot{Other choices of 
vacuum have been discussed in \refs{\gmsii,\nati}
and it was argued in \senissue\ that this redundancy in the choice of
vacuum reflects a bad choice of coordinates in field space. In any event,
the solution generating technique works best if we start from something
rather than nothing.}
\eqn\threek{\phi=1, \qquad C= \bar a, \qquad A_\mu = 0, \qquad \mu=0,1,
 \cdots 23.}

Although we do not know the form of the infinite number of
higher derivative terms in the effective action, we do know that
they transform covariantly under gauge transformations, and this is
sufficient to construct a solution using the solution generating
technique. Thus the transformation
\eqn\threel{\eqalign{\phi & \rightarrow U \phi \bar U \cr
                        C & \rightarrow U C \bar U \cr
                        A_\mu & \rightarrow U A_\mu \bar U \cr }}
generates a new solution from the vacuum as long as $\bar U U=1$ and
$U \bar U = 1-P_n$. Choosing $U=S^n$ then gives the solution
\eqn\threem{\eqalign{\phi & = S^n \bar S^n = (1-P_n) \cr
                       C & = S^n \bar a \bar S^n \cr
                       A_\mu & = 0. \cr }}

Although we know this is a solution which is localized in $\R^2$ and
hence represents a $23$-brane, it might seem difficult to calculate
the tension and spectrum of fluctuations about the solution without
detailed knowledge of the higher derivative terms in \threei. Luckily,
this is not the case and we can identify this solution with the
$D23$-brane of the bosonic string by computing the tension and
spectrum of low-lying fluctutations and showing that they agree
with those of a $D23$-brane.

To compute the tension note that $D \phi= \bar D \phi=0$ since
$D \phi$ vanishes in vacuum and transforms covariantly under the
solution generating transformation. Similarly, $DF= \bar D F=0$.
Since derivatives of $\phi$ and $F$ must appear as covariant
derivatives in the action, this shows that any term containing
derivatives of $\phi$ or $F$ does not contribute to the tension.
This leaves the potential term and terms involving $F^2$ and higher
powers of $F$ (i.e. those appearing in the expansion of the
DBI action).

Terms involving the gauge field strength also do not contribute to
the tension as a result of the vanishing of $V(\phi-1)$ in the closed
string vacuum. Specifically,  using $V(0)=0$ we find
\eqn\threen{V(\phi-1)[C,\bar C]^2 = V(-P_n)(1-P_n)=V(-1)P_n(1-P_n)=0.}
Thus the $V(\phi-1)(F+\Phi)^2$ term does not contribute to the tension.
Similarly, the  higher powers of
$(F+ \Phi)$ in an expansion of \threej\ are also multiplied by $V(\phi-1)$ 
and so  do
not contribute.

We are then left with the potential contribution to the tension.
Using
\eqn\threeo{V(\phi-1)=V(-P_n)=V(-1)P_n=P_n}
and
\eqn\threep{{\sqrt{G} \theta \over G_s} = {2 \pi \alpha' \over g_s}}
we find that the action evaluated for the solution \threem\ is
\eqn\threeq{S = (2 \pi)^2 \alpha' n {c \over g_s} \int d^{24} x.}
Thus the tension of the solution is
\eqn\threer{T_{23} = {(2 \pi)^2 \alpha' n c \over g_s} = (2 \pi)^2 \alpha'
n T_{25}^D = n T_{23}^D, }
with $T_p^D$ the tension of a $Dp$-brane of the bosonic string
as give in \pol.

Let me make a few comments concerning this solution and the
identification with a $D23$-brane:
\item{1.}
In addition to the tension matching that of a $D23$-brane, it is not
hard to 
show that the tachyon and gauge fields on the $D23$-brane arise
from a collective coordinate analysis of fluctuations about the solution.
For example, the solution preserves a $U(n)$ subgroup of $U({\cal H})$
acting in the image of $P_n$ and as in the usual Higgs mechanism this
leads to massless $U(n)$ gauge fields. This non-Abelian gauge symmetry
would be very difficult to see in a description of D-branes as
solitons in a commutative theory. An analysis of the tachyon field
fluctuations leads to tachyons in the adjoint of $U(n)$ on the
$D23$-brane \hklm.  One also finds massive fluctuations, but
their mass cannot be computed reliably without knowledge of the higher
derivative terms in \threei.
\item{2.} The discussion presented here (following \hkl) improves on
the discussion in \hklm\ by not requiring that one take the limit
$\alpha' B \rightarrow \infty$. In particular, the construction of
$D$-branes as noncommutative solitons holds for finite noncommutativity
and finite coupling constant.
\item{3.} The solution we have found has a non-zero gauge field. This
is to be contrasted with the constructions of $D$-branes as lumps in
truncated string field theory \tsolrefs\ or in background independent
string field theory where the gauge field vanishes. There is no obvious
contradiction since the transformation between the gauge field
variables here and those used in these other two treatments is
undoubtedly subtle and complicated. Still, it would be nice to
understand this better.
\item{4.}The construction can easily be extended to all $Dp$-branes with
$p$ odd by turning on a $B$ field in the even-dimensional transverse space
to the $Dp$-brane and repeating the above analysis with the obvious
modifications.

\subsec{Type II D-branes as Noncommutative Solitons}
These ideas can also be applied to construct $D$-branes in type II string
theory starting from unstable brane configurations. There are two
general classes of unstable brane configurations of interest. The first
consists of non-BPS $Dp$-branes for $p$ odd in IIA theory or $p$ even
in IIB theory. The second is the $Dp - \overline{Dp}$ system of BPS
branes and anti-branes for $p$ even in IIA and $p$ odd in IIB.
The first case is very similar to the analysis presented above for the
bosonic string \foot{Except for the subtle issue of how to interpret
the new tensionless brane solutions found in \refs{\dasgupta,\hklm}. In
\hklI\ it is proposed that these are gauge equivalent to the vacuum
solution.}. Because of this  I will focus on the
second case.

Consider a space-filling $D9- \overline{D9}$ system in IIB string theory.
The low-lying excitations consist of two gauge fields $A^{\pm}_\mu$ coming
from open strings which begin and end on the same  $D$-brane and a complex
tachyon $\phi$ from open strings which begin on one $D$-brane and end on the
other (the tachyon is complex because of the two orientations of these
open strings). This system is closely related to the field theory studied
at the beginning of sec 3.4. 
 The action of $A^\pm_\mu, \phi$ in type II string theory  has not been as well
studied as in the bosonic string, so we will have to proceed with a bit
more guesswork that in the previous analysis.

The action is constrained by  the presence of a $Z_2$ symmetry,
denoted by $(-1)^{F_L}$, which exchanges the $D9$ and $\overline{D9}$-branes
and takes
\eqn\threes{\eqalign{\phi & \leftrightarrow \bar \phi, \cr
                     A^+ & \leftrightarrow A^-~.\cr }}
Sen's conjectures regarding tachyon condensation and explicit calculations
referred to earlier in truncated open string field theory and BSFT
suggest a
potential of the form $V(\phi \bar \phi -1) + V(\bar \phi \phi -1)$
with $V(0)=1$ at the local maximum and a ring of minima at $|\phi|=1$
with $V(0)=0$.

In the noncommutative
theory $F^-$ and $F^+$ transform differently under the two $U(1)$ 
factors, but one can form linear combinations by noting that 
$F^-$ and $\phib F^+ \phi$ transform in the same way.  So an acceptable 
gauge kinetic term is 
\eqn\zgc{\eqalign{{\cal L}_{gauge} 
=&~ h_+(\phib \phi-1)\left\{F^-_{\mu\nu}+ \Phi_{\mu\nu}
+\phib( F^+_{\mu\nu}+ \Phi_{\mu\nu})\phi \right\}^2  \cr
+&~ h_-(\phib \phi-1)\left\{F^-_{\mu\nu}+ \Phi_{\mu\nu}
-\phib( F^+_{\mu\nu}+ \Phi_{\mu\nu})\phi \right\}^2 \cr
 +&~
 \left\{\phi \leftrightarrow \phib, \quad A^+ \leftrightarrow A^-\right\}.}}
The last line is included to enforce symmetry under $(-)^{F_L}$.  A similar
expression holds for terms including higher powers of the gauge fields.
We expect that the kinetic term for the ``center of mass'' $U(1)$
should vanish in the closed string vacuum which implies $h_+(0)=0$.
It is less clear whether  $h_-(0) =0$ or not. We will see that the
solution has the correct tension whether or not $h_-(0)=0$.

Tachyon kinetic terms appear as in \twocc, but now multiplied by functions
$f(\phib \phi-1)$ and symmetrized.  Symmetry under $(-)^{F_L}$ 
implies that the potential is
of the form
\eqn\zgd{V(\phi\phib-1)+V(\phib \phi-1)~.}
In the notation of \twocc\ we take $V=W$.

Now we use our solution generating transformation to construct exact
solitons representing BPS D-branes.  As before, we will explicitly
consider the codimension two case; starting with a spacefilling
$D9-\overline{D9}$ system of IIB this will produce  BPS $D7$-branes.
As in \twodd, the solution we generate starting from the closed string
vacuum is
\eqn\vorantvorb{\eqalign{
\phi =&~ S^{m} \bar{S}^{n}~,\cr
C^{-} =&~ S^{n}\ad \bar{S}^{n}~,\cr
C^{+} =&~ S^{m}\ad \bar{S}^{m}~, \cr
A^+_\mu =&~ A^-_\mu =0~, \quad \mu =0 \ldots 7~.
}}
We will see that this solution represents $m$ $D7$-branes coincident
with $n$ $\overline{D7}$-branes. 

We now work out the energy of this solution.  As in the bosonic case,
covariant derivatives of the tachyon and gauge field strengths 
vanish before, and thus  after the
solution generating 
transformation, and so do not contribute to the energy.  It is less
trivial to verify that the gauge field terms \zgc\ do not contribute.
We need to compute
\eqn\zgd{h_+(\phib \phi -1)
\left\{[C^-,\overline{C^-}] + \phib[C^+,\overline{C^+}]\phi
\right\}^2 +
h_-(\phib \phi-1)\left\{[C^-,\overline{C^-}] - \phib [C^+,\overline{C^+}]\phi
\right\}^2.}
For the  solution,
\eqn\zge{\eqalign{\phi \phib =&~ I-P_m~, \cr
\phib \phi =&~  I-P_n~, \cr
[C^-, \overline{C^-}]=&~-(I-P_n)~,\cr
 \phib[C^+, \overline{C^+}]\phi=&~-(I-P_n)~,}}
the first term in \zgd\ vanishes since $h_+(\phib \phi-1)=h_+(-P_n)
=h_+(-1)P_n$, which is orthogonal to $I-P_n$;  here we used that
$h_+(0)=0$.   The second term in \zgd\ vanishes without a similar assumption
about $h_-$.   So, as in the bosonic theory, the only contribution to
the energy comes from the potential term, which we find to be 
\eqn\zgf{V(\phi\phib-1)+V(\phib \phi-1) = V(-P_m)+V(-P_n) = V(-1)(P_m +P_n).}
Repeating the computation leading to \threer\ in the bosonic case now
gives the tension
\eqn\zgg{T_{nm} = {(2\pi)^2 \alpha' (n+m) c \over g_s}T_9^D = (n+m)T_7^D.}
as expected for $m$ $D7$-branes plus $n$ $\overline{D7}$-branes.

Using the ABS construction of solitons in section 3.4, it is 
straightforward to generalize the above discussion to codimension
$2p$ solitons representing coincident $D(9-2p)$ and 
$\overline{D(9-2p)}$ branes. It is also possible to extend the previous
discussion of the fluctuation spectrum to show that the expected low-lying
spectrum of excitations ($U(n) \times U(m)$ gauge fields and tachyons in
the $(n,m)$) arise on the lower-dimensional $D$-branes.

\newsec{Lecture 4}

\subsec{Noncommutative Solitons and $K$-theory}
In the previous lectures we have developed an approach to $D$-branes that
describes them as solitons in noncommutative field theory. We have seen
that the tension and the low-lying spectrum of excitations are both
correctly obtained in this picture. It is natural to ask whether other
aspects of $D$-branes can be understood from this point of view or whether
new points of view on $D$-brane physics are suggested by the noncommutative
approach. Since $B=0$ is a special value while $B \ne 0$ is generic, one
might argue that the noncommutative description is more general, and hence
the proper formulation of $D$-branes should naturally involve concepts
of noncommutative geometry.

Until recently field theorists have been able to rely on the basic tools
of algebraic topology (that is homotopy, homology and cohomology) to compute
the charges of solitons. Roughly speaking, these tools (or functors) provide
a natural way to associate an Abelian group (representing the charges) to
a topological space. When it comes to $D$-branes new tools are needed because
$D$-branes naturally carry gauge fields and thus  one needs a way to
associate an Abelian group to topological spaces equipped with vector bundles.
This leads to the identification of  $D$-brane 
charge with  $K$-theory
\refs{\mm,\ewk,\phk}. The
noncommutative description of $D$-branes as solitons  makes this fact manifest,
but also suggests some
interesting connections to some more exotic  aspects of K-theory and 
noncommutative geometry \refs{\wittenstrings,\matsuo,\harmor}. 
The correct tool for studying
$D$-brane charge in noncommutative field theory  might be
called noncommutative algebraic topology and should associate Abelian groups
to  $C^*$ operator algebras. This leads to the $K$-theory of $C^*$ algebras.
For reasons of both time and competence I will only discuss some elementary
aspects of this theory. A brief summary of some of the relevant mathematics
can be found in \msinger. For more details
see \refs{\wegge,\blackadar,\kadison}.

To start with let me a give a low-brow summary of a construction which is
central to K theory.
An Abelian semi-group is a set $S={a,b,c,\cdots}$ with  a binary
Abelian composition law $\circ$ 
\eqn\foura{a \circ b = b \circ a = c \in S~.}
We do not  assume the existence of an inverse or an identity, hence this
is a semi-group rather than a group. Associated to $S$ there is a
Grothendieck-given Abelian group, $G(S)$, which consists of ordered pairs
of elements of $S$, $(a,b)$ with the identification
\eqn\fourb{(a,b) \sim (a \circ c, b \circ c), \qquad c \in S~.}
A group multiplication law, $+$, can be defined for these ordered pairs by
the rule
\eqn\fourc{(a,b) + (c,d) \equiv (a \circ c, b \circ d)~.}
It is simple to check that this defines an Abelian group with
identity ${\rm id} = (a,a)$ for any $a \in S$ and with inverse
$(a,b)^{-1} = (b,a)$:
\eqn\fourd{\eqalign{(a,a)+(c,d) & = (a \circ c, a \circ d) =
(c \circ a, d \circ a) = (c,d) \cr
                     (a,b)+(b,a) & = (a \circ b,b \circ a) =
                     (a \circ b,a \circ b) = {\rm id}~. \cr }}

We now consider some examples of this construction.
\item{1.}Take the set $S$ to consist of the set of natural numbers,
$S=N={1,2,3, \cdots}$ and the composition law $\circ$ to be
addition. Then it is easy to check that $G(S)$ is isomorphic
to the Abelian group formed by the set of all integers with
group multiplication being addition of integers and
$n^{-1}= -n$. The isomorphism identifies the ordered pair $(n,m)$
with the integer $n-m$.
\item{2.}Take the set $S$ to consist of  vector bundles over
a manifold $X$. The composition law $\circ$ is $\oplus$, the direct
sum of vector bundles. Thus we identify the pair of vector bundles
$(E,F)$ with $(E \oplus G, F \oplus G)$ for an arbitrary vector
bundle $G$. The K group of $X$ (or $K_0(X)$) is the Grothendieck
group $G(S)$. In \absapp\ the above equivalence relation was
interpreted as the creation of brane-antibrane pairs following
the work of Sen.
\item{3.}Let $S=[X,{\cal F}]$ be the set of homotopy classes of maps
from a compact manifold $X$  into the space of Fredholm operators
on Hilbert space \foot{Recall that an operator $O$ is Fredholm if its 
image is closed and its
kernel and cokernel are both finite dimensional}. It is proved in \atiyah\  
that one can define a composition law such that
once again $G(S)=K(X)$. If $X$ is a single point then the set of
homotopy classes of maps from $X$ into ${\cal F}$ has one element for
each disconnected component of ${\cal F}$. Since the disconnected components
are labelled by the index, this shows that 
$K({\rm point})=Z$.
\item{4.}Let $S$ be the set of Murray von-Neumann equivalence classes of
projection operators in a $C^*$ algebra ${\cal A}$. Recall that 
a $C^*$ algebra  can be thought of as a self-adjoint
subalgebra of the algebra of bounded operators on Hilbert space, $B({\cal H})$.
Two projection operators $p$, $q$ are Murray-von Neumann equivalent if
$p= \bar v v$, $q= \bar v v$ for some partial isometry $v$. In general the
sum of two projection operators is not a projection operator, so to define
a composition law one must ``stabilize'' by considering infinite dimensional
matrices with entries in ${\cal A}$. One can then move projection operators
down the diagonal to make them orthogonal and define a 
binary composition law on $S$.   The
Abelian group $G(S)$ constructed this way 
is called the K group of  ${\cal A}$, $K({\cal A})$.
As an example, the space of continuous functions on a manifold $X$ is
a commutative $C^*$ algebra, $C(X)$ with a norm defined by
$||f||= {\rm sup}_{x \in X} f(x)$.  It is a standard result \wegge\ that
$K(C(X))=K(X)$.

The second definition of the $K$-theory group of a manifold $X$ has
been discussed extensively in connection with $D$-branes. The third and
fourth definitions on the other hand involve concepts that we have
already encountered in the description of $D$-branes as noncommutative
solitons \refs{\matsuo,\harmor}. 
For example, consider tachyon condensation on a $D9-
\overline{D9}$ system in $IIB$ string theory to make a $D7$-brane as
described in the previous lecture. In order to have finite action
the tachyon field $\phi$ must be Fredholm and we saw that the induced
$D7$-brane charge was just the index of $\phi$. Let $X$ denote the
$D7$-brane world volume and consider $X$-dependent tachyon configurations.
Finite action configurations of this sort give us a map
\eqn\fredmap{T: X \rightarrow {\cal F}}
of $X$ into Fredholm operators. These configurations will be classified
by the homotopy class of the map \fredmap, which according to the
third definition is just $K(X)$. We could also consider such a construction
in IIA theory starting with an unstable $D9$-brane. In this case
the tachyon field is real, and so we should consider maps of $X$
(the world volume of a lower-dimensional $D$-brane constructed as
a noncommutative soliton) into the space of self-adjoint Fredholm operators,
${\cal F}^{sa}$. 
The homotopy classes of maps $[X,{\cal F}^{sa}]$ provides a model of
the K-group 
$K^1(X)$ \atiyahs, an identification which was utilized in
the proposal of \wittenstrings.

\subsec{$D$-brane Charge and Toeplitz Operators}
There is a deeper aspect of the connection between noncommutative
tachyon condensation and $K$-theory which I would like to outline
briefly. To begin with, let me consider the related problem of
understanding the connection between the topological charge and the
index of Fredholm operators as it appears in the constructions
described in the previous lecture. For example, in the noncommutative
set-up the $D7$-brane charge is the index of the tachyon operator
$\phi$. On the other hand, in the commutative theory the $D7$-brane
charge is given by the winding number of the classical tachyon field,
\eqn\foure{Q_{D7} \sim {1 \over 2 \pi |\phi(\infty)|^2}
 \int_{S^1_\infty} \phi d \phi \sim {1 \over 2 \pi} \int_{R^2} F~. }
What is the connection between these two facts? We need an operator
analog of the winding number. To construct this we consider a model
for the shift operator $S$ which appears in the construction of the
$D7$-brane solution where $S$ acts on the Hilbert space of
states on a circle (which we think of physically as an $S^1$
encircling the $D7$-brane).

So consider ${\cal H}=L^2(S^1)$ with orthonormal basis
\eqn\fourf{\psi_n(\theta)=
{e^{i n \theta} \over \sqrt{2 \pi}}, \qquad n \in Z~. }
A function in the $C^*$ algebra of functions on $S^1$, $f \in C(S^1)$
determines an operator $M_f$ acting on ${\cal H}$ in the obvious way:
\eqn\fourg{M_f \psi(\theta) = f(\theta) \psi(\theta)~.}
An operator which shifts $\psi_n \rightarrow \psi_{n+1}$ has index
$0$ since now $n$ runs over all integers. To find an analog of the
shift operator we had in the harmonic oscillator basis for $L^2(R)$
we define a subspace ${\cal H}_+$ of ${\cal H}$ by
\eqn\fourh{{\cal H}_+ = {\rm span} \{ \psi_n(\theta), ~n \ge 0 \} }
and let $P$ be the projection operator onto ${\cal H}_+$. We can
think of $H_+$ as the space of boundary values of holomorphic functions
on $\R^2$.

We now define a Toeplitz  operator on ${\cal H}_+$ which generalizes the
shift operator by multiplying by a general $f(\theta)$ and then
projecting back to ${\cal H}_+$:
\eqn\fouri{T_f= P M_f :~ {\cal H}_+ \rightarrow {\cal H}_+~. }
We can see that this generalizes the shift operator by noting that
the Toeplitz operator constructed from $f_l(\theta)=e^{i \l \theta}$
for $l>0$ acts as the $l^{\rm th}$ power of the shift operator:
\eqn\fourj{T_{f_l}:~ \psi_n(\theta) \rightarrow \psi_{n+l}(\theta)~.}
An index theorem for Toeplitz operators \countb\ says that
\eqn\fourk{{\rm ind}~ T_f = {1 \over 2 \pi} \int \bar f d f }
thus providing the desired connection between the index and winding
number. This argument and the index theorem can be extended to
include the noncommutative ABS configurations defined in $\R^{2n}$ \harmor.

\subsec{BDF and all that}
The formalism described above provides one of the simplest examples
of a general structure analyzed by Brown, Douglas and Fillmore (BDF) \bdf\
which may ultimately be of some importance in understanding D-branes.
To explain this, note that 
\eqn\fourl{T_f T_g - T_{fg} \in {\cal K({\cal H})}}
where $ {\cal K({\cal H})}$ is the space of compact operators.
For example,
\eqn\fourm{T_{f_l} T_{f^*_l}-T_1 = P_l}
where $P_l$ is the projection operator onto the space spanned by
$\psi_0,\psi_1, \cdots \psi_{l-1}$.

The Toeplitz operators  form a $C^*$ algebra, which
we  denote by  ${\cal T}$,  which maps
to the $C^*$ algebra $C(S^1)$ of continuous  functions on $S^1$ by
$T_f \rightarrow f$. Furthermore, since $T_f T_g - T_{fg}$ is compact
and $fg-gf=0$ in $C(S^1)$ (since $C(S^1)$ is commutative),
the kernel of this map consists of the
compact operators.  We have therefore deduced the existence of a
short exact sequence
\eqn\fourn{0 \rightarrow {\cal K} \rightarrow {\cal T} \rightarrow
C(S^1) \rightarrow 0.}

Exact sequences of this form  play an important role in the
study of operator algebras and their invariants. In particular,
BDF classified the ``extensions of $C(X)$ by ${\cal K}$''
meaning the possible $C^*$ algebras ${\cal A}$ such that one has
a short exact sequence
\eqn\fouro{0 \rightarrow {\cal K} \rightarrow {\cal A} 
\rightarrow C(X) \rightarrow
0~.}
This classification is based on an invariant of such extensions called
the Busby invariant and an associated Abelian group, 
${\bf Ext}(C(X),{\cal K})$.
Without going into detail, it is possible to motivate the definitions
of these objects by using the Toeplitz algebra as an example.

For solitons in commutative field theory the topology is captured by
the gauge invariant ``winding'' of the fields at infinity.  We 
have seen that  the Weyl transform maps bounded operators to functions
which can be non-vanishing at infinity. Compact operators on the other
hand map to functions vanishing at infinity. Thus we expect the 
``topology'' of a noncommutative field theory to be independent of 
a change by compact operators, that is the topology should depend only
on  the  quotient
of bounded operators by compact operators, $Q({\cal H}) = B({\cal H})/
{\cal K}({\cal H})$, otherwise known as the Calkin algebra. 

The Busby invariant is a map from $C(X)$ into $Q({\cal H})$ defined
for $f \in C(X)$ by choosing a $T_f \in {\cal A}$ which maps to $f$
and defining $\tau(f) = \pi(T_f)$ where $\pi: B{(\cal H}) \rightarrow
Q({\cal H})$ is the projection. Two configurations should have the
same topology if they are gauge equivalent, in operator language
this is known as ``strong equivalence''. Two extensions are strongly
equivalent if the Busby invariants are related by
$\tau_2(f) = \pi(U) \tau_1(f) \pi(U)^*$. The set of strong equivalence
classes of extensions of $C(X)$ by ${\cal K}$ is denoted 
${\bf Ext}(C(X),{\cal K})$. It is possible to show that one can define
a sum of extensions that turns ${\bf Ext}(C(X),{\cal K})$ into an
Abelian group which we can identify with the $D$-brane charge.
These strong equivalence classes of extensions can be used to define
a variant of K-theory sometimes known as K-homology.
This and other
considerations discussed in \refs{\periwal,\harmor}
suggest that $D$-brane charge
should really be associated to $K$-homology.
Some other aspects of $K$-theory such as Bott periodicity also find a
natural setting in the language of noncommutative solitons.

In these lectures I have only described D-branes as noncommutative
solitons in $\R^{2n}$. It is obviously interesting to extend these
constructions to more complicated spaces. See \refs{\bars,\emilgreg} 
and references
therein for the extension to tori and orbifolds. 

One motivation for the study of tachyon condensation in open string theory
is the hope that open string theory might provide an alternate starting
point for a fundamental formulation of string theory. If this is correct
and amenable to practical analysis, then closed fundamental strings and
NS-branes must also make an appearance. Closed strings of course appear
in perturbation theory about the unstable vacuum, but finding them in the
stable vacuum in terms of open string fields is problematic. 
Closed strings and NS-branes do not have the correct
tension to be constructed as solitons of open string field theory, at least
naively. There have been attempts to describe closed fundamental strings
as electric flux tubes
in the tachyon condensed vacuum \refs{\yi,\bhy,\hklm,\ghy,\senfun,\kls,
\shatstring} 
but we seem far from a definitive
picture. The description of NS-branes seems even more difficult since they
are naturally viewed as solitons of closed string field theory \refs{\hetsol,
\chsa,\chsb,\chsc}. A preliminary attempt has been made to describe $D$-branes
as noncommutative solitons in the presence of NS-branes by using twisted
$U({\cal H})/U(1)$ bundles \harmor. It would be interesting to pursue these
ideas further.

\bigskip\medskip\noindent
{\bf Acknowledgements:}
I would like to thank the organizers for the invitation to present
these results at the Komaba 2000 meeting and for their hospitality
during my visit. I am greatly indebted to
Per Kraus, Finn Larsen,  Emil Martinec and Greg Moore
for their collaboration on
the topics presented here. I thank Greg Moore and 
Li-Sheng Tseng  for
comments on  the manuscript. 
In addition I would like to thank Rouyu Bao,
Steuard Jensen, David Kutasov, Nati Seiberg, Andy
Strominger, and Cosmas Zachos
for discussions of various aspects of noncommutative field theory.
This work was supported in part by NSF grant PHY-9901194.

\listrefs
\bye